\documentclass[apj]{emulateapj}
\newcommand\submitms{n}		
\if\submitms y
\else
\usepackage{apjfonts}
\fi

\usepackage{ifthen}
\usepackage{natbib}
\usepackage{amssymb, amsmath}
\usepackage{appendix}

\if\submitms y
\bibpunct[, ]{(}{)}{,}{a}{}{,}
\else
\fi

\shorttitle{Thermal Emission of WASP-14\MakeLowercase{b}}
\shortauthors{Blecic {\em et al.}}

\newcommand\degree{\degr}
\newcommand\degrees\degree

\DeclareSymbolFont{UPM}{U}{eur}{m}{n}
\DeclareMathSymbol{\umu}{0}{UPM}{"16}
\let\oldumu=\umu
\renewcommand\umu{\ifmmode\oldumu\else\math{\oldumu}\fi}
\newcommand\micro{\umu}
\renewcommand\micron{\micro m}
\newcommand\microns \micron

\let\oldsim=\sim
\renewcommand\sim{\ifmmode\oldsim\else\math{\oldsim}\fi}
\let\oldpm=\pm
\renewcommand\pm{\ifmmode\oldpm\else\math{\oldpm}\fi}
\newcommand\by{\ifmmode\times\else\math{\times}\fi}
\newcommand\ttt[1]{10\sp{#1}}

\newcommand\tablebox[1]{\begin{tabular}[t]{@{}l@{}}#1\end{tabular}}
\newbox{\wdbox}
\renewcommand\c{\setbox\wdbox=\hbox{,}\hspace{\wd\wdbox}}
\renewcommand\i{\setbox\wdbox=\hbox{i}\hspace{\wd\wdbox}}

\newcount\timect
\newcount\hourct
\newcount\minct
\newcommand\now{\timect=\time \divide\timect by 60
         \hourct=\timect \multiply\hourct by 60
         \minct=\time \advance\minct by -\hourct
         \number\timect:\ifnum \minct < 10 0\fi\number\minct}

\newcommand\mctc{\multicolumn{2}{c}}

\catcode`@=11

\if\submitms y
\renewcommand\comment[1]{}
\else
\newcommand\comment[1]{}
\fi
\newcommand\commenton{\catcode`\%=14}
\newcommand\commentoff{\catcode`\%=12}

\renewcommand\math[1]{$#1$}
\newcommand\mathshifton{\catcode`\$=3}
\newcommand\mathshiftoff{\catcode`\$=12}

\comment{the backslash is necessary}

\comment{alignment tab}

\let\atab=&
\newcommand\atabon{\catcode`\&=4}
\newcommand\ataboff{\catcode`\&=12}

\let\oldmsp=\sp
\let\oldmsb=\sb
\def\sp#1{\ifmmode
           \oldmsp{#1}%
         \else\strut\raise.85ex\hbox{\scriptsize #1}\fi}
\def\sb#1{\ifmmode
           \oldmsb{#1}%
         \else\strut\raise-.54ex\hbox{\scriptsize #1}\fi}
\newbox\@sp
\newbox\@sb
\def\sbp#1#2{\ifmmode%
           \oldmsb{#1}\oldmsp{#2}%
         \else
           \setbox\@sb=\hbox{\sb{#1}}%
           \setbox\@sp=\hbox{\sp{#2}}%
           \rlap{\copy\@sb}\copy\@sp
           \ifdim \wd\@sb >\wd\@sp
             \hskip -\wd\@sp \hskip \wd\@sb
           \fi
        \fi}
\def\msp#1{\ifmmode
           \oldmsp{#1}
         \else \math{\oldmsp{#1}}\fi}
\def\msb#1{\ifmmode
           \oldmsb{#1}
         \else \math{\oldmsb{#1}}\fi}
\def\supon{\catcode`\^=7}
\def\supoff{\catcode`\^=12}
\def\subon{\catcode`\_=8}
\def\suboff{\catcode`\_=12}
\def\supsubon{\supon \subon}
\def\supsuboff{\supoff \suboff}

\comment{
\let\oldmsp=\sp
\let\oldmsb=\sb
\renewcommand\sp[1]{\ifmmode
	   \oldmsp{#1}%
	 \else\strut\raise.85ex\hbox{\scriptsize #1}\fi}
\renewcommand\sb[1]{\ifmmode
	   \oldmsb{#1}%
	 \else\strut\raise-.54ex\hbox{\scriptsize #1}\fi}
\newcommand\msp[1]{\ifmmode
	   \oldmsp{#1}
	 \else \math{\oldmsp{#1}}\fi}
\newcommand\msb[1]{\ifmmode
	   \oldmsb{#1}
	 \else \math{\oldmsb{#1}}\fi}
\newcommand\supon{\catcode`\^=7}
\newcommand\supoff{\catcode`\^=12}
\newcommand\subon{\catcode`\_=8}
\newcommand\suboff{\catcode`\_=12}
\newcommand\supsubon{\supon \subon}
\newcommand\supsuboff{\supoff \suboff}
}

\newcommand\actcharon{\catcode`\~=13}
\newcommand\actcharoff{\catcode`\~=12}

\newcommand\paramon{\catcode`\#=6}
\newcommand\paramoff{\catcode`\#=12}

\comment{And now to turn us totally on and off...}

\newcommand\reservedcharson{\commenton \mathshifton \atabon \supsubon \actcharon
	\paramon}

\newcommand\reservedcharsoff{\commentoff \mathshiftoff \ataboff
	\supsuboff \actcharoff \paramoff}

\catcode`@=12
\reservedcharsoff

\reservedcharson

\comment{ Must have ONLY ONE of these... trust these macros, they work

}
\reservedcharsoff

\actcharon
\if\submitms y

\else

\comment{
\received{}
\revised{}
\accepted{}
}
\fi

\if\submitms n
\slugcomment{\tablebox{
In preparation for {\em ApJ}. \\
DRAFT of {\today}.}}
\else
\slugcomment{Published 2013 November 15 in ApJ}
\journalinfo{}
\fi

\reservedcharson
\begin{document}


\slugcomment{Published 2013 November 15 in ApJ}

\title {Thermal Emission of WASP-14\MakeLowercase{b} Revealed with Three {\em Spitzer} Eclipses}

\author{Jasmina Blecic\altaffilmark{1},      Joseph Harrington\altaffilmark{1, 2},
        Nikku Madhusudhan\altaffilmark{3},   Kevin B.\ Stevenson\altaffilmark{1},
        Ryan A.\ Hardy\altaffilmark{1},      Patricio E. Cubillos\altaffilmark{1, 2}, 
        Matthew Hardin\altaffilmark{1},     Christopher J.\ Campo\altaffilmark{1}, 
        William C.\ Bowman\altaffilmark{1},  Sarah Nymeyer\altaffilmark{1},
        Thomas J.\ Loredo\altaffilmark{4},        
        David R.\ Anderson\altaffilmark{5},  and Pierre F.\ L.\ Maxted\altaffilmark{5}}

\affil{\sp1 Planetary Sciences Group, Department of Physics, University of Central Florida, Orlando, FL 32816-2385, USA}

\affil{\sp2 Max-Planck-Institut f\"{u}r Astronomie, D-69117 Heidelberg, Germany}

\affil{\sp3 Department of Physics and Department of Astronomy, Yale University, New Haven, CT 06511, USA}

\affil{\sp4 Center for Radiophysics and Space Research, Space Sciences Building, Cornell University Ithaca, NY 14853-6801, USA}

\affil{\sp5 Astrophysics Group, Keele University, Keele, Staffordshire ST5 5BG, UK}

\email{jasmina@physics.ucf.edu}

\begin{abstract}

Exoplanet WASP-14b is a highly irradiated, transiting hot Jupiter. Joshi et al. calculate an equilibrium temperature (\math{T\sb{\rm eq}}) of 1866 K for zero albedo and reemission from the entire planet, a mass of 7.3 {\pm} 0.5 Jupiter masses (\math{M\sb{\rm J}}) and a radius of 1.28 {\pm} 0.08 Jupiter radii (\math{R\sb{\rm J}}). Its mean density of 4.6 g\,cm\sp{-3} is one of the highest known for planets with periods less than 3 days. We obtained three secondary eclipse light curves with the {\em Spitzer Space Telescope}. The eclipse depths from the best jointly fit model are 0.224\% {\pm} 0.018\% at 4.5 {\micron} and 0.181\% {\pm} 0.022\% at 8.0 {\micron}. The corresponding brightness temperatures are 2212 {\pm} 94 K and 1590 {\pm} 116 K. A slight ambiguity between systematic models suggests a conservative 3.6 {\micron} eclipse depth of 0.19\% {\pm} 0.01\% and brightness temperature of 2242 {\pm} 55 K. Although extremely irradiated, WASP-14b does not show any distinct evidence of a thermal inversion. In addition, the present data nominally favor models with day night energy redistribution less than \sim30\%. The current data are generally consistent with oxygen-rich as well as carbon-rich compositions, although an oxygen-rich composition provides a marginally better fit. We confirm a significant eccentricity of \math{e} = 0.087 {\pm} 0.002 and refine other orbital parameters.







\if\submitms y
\else
\fi
\end{abstract}


\section{INTRODUCTION}
\label{intro}

The {\em Spitzer Space Telescope} \citep{Werner2004} is the most widely used facility for measuring thermal properties of extrasolar planets. {\em Spitzer} systematics are well studied and modeled, providing an invaluable resource for exoplanet characterization \citep{SeagerDeming2010}. This has enabled the measurement of tens of atmospheres, using the detection of primary and secondary eclipses as the most prolific method of investigation to date.

The planet-to-star flux ratio is enhanced in the infrared due to the rising planetary thermal emission and the dropping stellar emission, enabling detection of planetary emission through high-precision photometric measurements. Combining several secondary-eclipse observations measured in broad {\em Spitzer} bandpasses with the Infrared Array Camera \citep[IRAC; ][]{Fazio2004IRAC}, a low-resolution dayside spectrum from the planet can be reconstructed, revealing key atmospheric and physical parameters. These measurements can further be used to constrain atmospheric composition, thermal structure, and ultimately the formation and evolution of the observed planet.

WASP-14b represents an intriguing object for such an analysis, having characteristics not so common for close-in, highly irradiated giant planets. \citet{Joshi2009-WASP14b} discovered it as a part of the SuperWASP survey \citep[Wide-Angle Search for Planets; ][]{Pollocco06, Cameron2006-SuperWASP, Cameron2007-SuperWASP}. Photometric and radial-velocity observations revealed a planetary mass of 7.3 {\pm} 0.5 \math{M\sb{\rm J}} and a radius of 1.28 {\pm} 0.08 \math{R\sb{\rm J}}. Its density (\math{\rho} = 4.6 g\,cm\sp{-3}) is significantly higher than typical hot-Jupiter densities of 0.34--1.34 g\,cm\sp{-3} \citep{Loeillet2008}. The planet is also very close to its star (semi-major axis 0.036 {\pm} 0.001 AU), and has a significant orbital eccentricity, refined slightly to \math{e} = 0.087 {\pm} 0.002 in this work.

Detailed spectroscopic analyses of the stellar atmosphere determined that the star belongs to the F5 main-sequence spectral type with a temperature of 6475 {\pm} 100 K and high lithium abundance of log \math{N}(Li) = 2.84 {\pm} 0.05. F-type stars with this temperature should have depleted Li, being close to the Li gap or ``Boesgaard gap'' \citep{BoesgaardTripicco1986, Balachandran1995}. However, the high amount of Li and a relatively high rotational speed of \math{v\sin(i)} = 4.9 {\pm} 1.0 km\,s\sp{-1} indicate that WASP-14 is a young star. Comparing these results with models by \citet{FortneyEtal2007apjPlanetRadii} for the range of planetary masses and radii led \citet{Joshi2009-WASP14b} to constrain the age of the system to 0.5--1.0 Gyr. 

\citet{Joshi2009-WASP14b} also discuss the high eccentricity of the planet. Because WASP-14b has a very small orbital distance, probable scenarios for such a significant eccentricity (their \math{e} = 0.091 {\pm} 0.003) would be either that the system age is comparable to the tidal circularization time scale or there is a perturbing body.

\citet{Husnoo2011-WASP14b} performed long-term radial-velocity measurements to discover or reject the presence of a third body. They refined the orbital eccentricity to \math{e} = 0.088 {\pm} 0.003.  They argue that this planet has undergone some degree of orbital evolution, but that it is still subject to strong tidal forces. They state that since there is no observable unambiguous trend in residuals with time, there is no firm evidence for a planetary companion. This would establish a new lower limit for the semimajor axis at which orbital eccentricity can survive tidal evolution for the age of the system.


\begin{table}[ht]
\vspace{-10pt}
\caption{\label{table:ObsDates} 
Observation Information}
\atabon\strut\hfill\begin{tabular}{lccc@{ }c@{ }c@{ }l}
    \hline
    \hline 
    Channel                     & Observation          & Start Time   & Duration      & Exposure           & Number of                   \\
                                & Date                 & (JD)        & (s)            & Time (s)           & Frames                      \\
    \hline
    \multicolumn{6}{c}{Main science observation}                                                                                         \\
    \hline 
    Ch1                         & 2010 Mar 18             & 2455274.4707   & 28055.4         & 2                     & 13760              \\
    Ch2                         & 2009 Mar 18             & 2454908.8139   & 19998.7         & 2 \math{\times} 2       & 2982               \\
    Ch4                         & 2009 Mar 18             & 2454908.8139   & 19998.7         & 12                    & 1481               \\ 
    \hline
    \multicolumn{6}{c}{Pre-observation}                                                                                                  \\ 
    \hline 
    Ch2+4                            & 2009 Mar 18              & 2454908.7877   & 2019            & 2                     & 213          \\ 
    \hline
    \multicolumn{6}{c}{Post-observation}                                                                                                 \\ 
    \hline
    Ch2+4                             & 2009 Mar 18             & 2454909.0455    & 367             & 2 \math{\times} 2,12  & 10            \\
   \hline
    
\end{tabular}\hfill\strut\ataboff
\vspace{-10pt}
\end{table}

We obtained three secondary eclipse light curves at 3.6 {\micron}, 4.5 {\micron}, and 8.0 {\micron} using {\em Spitzer}. We present analytic light-curve models that incorporate corrections for systematic effects that include the new \citet{StevensonEtal2012apjHD149026b} pixel sensitivity mapping technique, a Keplerian orbital model, estimates of infrared brightness temperatures, and constraints on atmospheric composition and thermal structure.

In Section \ref{sec:obs} we describe our observations. Section \ref{sec:data_red} discusses data reduction procedures. Section \ref{sec:Phot} presents our photometry and Section \ref{sec:SecEclResults} discusses the modeling techniques and results from each dataset. Section \ref{sec:orbit} presents constraints on the orbit of WASP-14b, and Section \ref{sec:atm} reveals the atmospheric structure and composition. In Section \ref{sec:discus} we discuss our results and in Section \ref{sec:concl} we present our conclusions. Data files containing the light curves, best-fit models, centering data, photometry, etc., are included as electronic supplements to this article.

\section{OBSERVATIONS}
\label{sec:obs}

The {\em Spitzer} IRAC instrument observed two events; one at 3.6 {\microns} in 2010 March (Knutson's program 60021, Warm {\em Spitzer}) and one observation simultaneously in two wavelength bands (4.5 and 8.0 {\microns}) in 2009 March (Harrington's program 50517, {\em Spitzer} cryogenic mission). The observation at 3.6 {\microns} (channel 1) was made in subarray mode with 2 s exposures, while the observations at 4.5 and 8.0 {\microns} (channels 2 and 4) were made in stellar mode (2\math{\times}2,12) with pairs of 2 s frames taken in the 4.5 {\micron} band for each 12 s frame in the 8.0 {\microns} band. This mode was used to avoid saturation in channel 2. 

We have pre- and post-observation calibration frames for the 4.5 and 8.0 {\microns} observation. Prior to the main observation, we exposed the array to a relatively bright source (see Section\ \ref{sec:ch2}). That quickly saturated charge traps in the detector material, reducing the systematic sensitivity increase during the main observation. Post-eclipse frames of blank sky permit a check for warm pixels in the aperture. The {\em Spitzer} pipeline version used for the 3.6 {\micron} observation is S.18.14.0 and for the 4.8 and 8.0 {\microns} observation is S18.7.0. The start date of each observation, duration, exposure time and total number of frames are given in Table\ \ref{table:ObsDates}. 

\section{DATA REDUCTION}
\label{sec:data_red}

\subsection{Background}
\label{sec:backgr}

Our analysis pipeline is called Photometry for Orbits, Eclipses and Transits (POET). It produces light curves from {\em Spitzer} Basic Calibrated Data (BCD) frames, fits models to the light curves, and assesses uncertainties. The derived parameters constrain separate orbital and atmospheric models. In this section we give a general overview of POET. Subsequent sections will provide details as needed.

Each analysis starts by identifying and flagging bad pixels in addition to the ones determined by the {\em Spitzer} bad pixel mask (see Section\ \ref{sec:Phot}). Then we perform centering. Due to the \sim 0.1\% relative flux level of secondary-eclipse observations and {\em Spitzer's} relative photometric accuracy of 2\% \citep{Fazio2004IRAC}, we apply a variety of centering routines, looking for the most consistent. We test three methods to determine the point-spread function (PSF) center precisely: center of light, two-dimensional Gaussian fitting, and least asymmetry (see Supplementary Information of \citealp{StevensonEtal2010Natur} and \citealp{LustEtal2013apjCentering}). The routines used for each data set are given below. We then apply 5\math{\times}-interpolated aperture photometry \citep{HarringtonEtal2007natHD149026b}, where each image is re-sampled using bilinear interpolation. This allows the inclusion of partial pixels, thus reducing pixelation noise \citep{StevensonEtal2012apjHD149026b}. We subtract the mean background within an annulus centered on the star and discard frames with bad pixels in the photometry aperture.

{\em Spitzer}\/ IRAC has two main systematics, which depend on time and the sub-pixel position of the center of the star.  To find the best time-dependent model (the ``ramp''), we fit a variety of systematic models from the literature, and some of our own, using a Levenberg--Marquardt \math{\chi\sp{2}} minimizer \citep{Levenberg1944, Marquardt1963}. We use our newly developed \citep{StevensonEtal2012apjHD149026b} BiLinearly Interpolated Subpixel Sensitivity (BLISS) mapping technique to model intrapixel sensitivity variation (see Section\ \ref{sec:sys}).  The BLISS method can resolve structures inaccessible to the widely used two-dimensional polynomial fit \citep{Knutson08, MachalekEtal2009ApJ-XO2b, FressinEtal2010ApJ-Tres3}.  It is faster and more accurate than the mapping technique developed by \citet{BallardEtalr2010PASP-NewIntrapixel}, which uses a Gaussian-weighted interpolation scheme and is not feasibly iterated in each step of Markov-Chain Monte Carlo (MCMC; see next section for details on modeling systematics).

To determine the best aperture size, we seek the smallest standard deviation of normalized residuals (SDNR) among different aperture sizes for the same systematic model components. The best ramp model at that aperture size is then determined by applying the Bayesian (BIC) and Akaike (AIC) information criteria \citep{Liddle2008}, which compare models with different numbers of free parameters (see Section \ref{sec:LightCurves}). The BIC and AIC cannot be used to compare BLISS maps with differing grid resolutions, or BLISS versus polynomial maps (see Section \ref{sec:sys}), but BLISS has its own method for optimizing its grid \citep{StevensonEtal2012apjHD149026b}.

To explore the parameter space and to estimate uncertainties, we use an MCMC routine (see Section \ref{sec:LightCurves}).  We model the systematics and the eclipse event simultaneously, running four independent chains until the \citet{GelmanRubin1992} convergence test for all free parameters drops below 1\%. Our MCMC routine can model events separately or simultaneously, sharing parameters such as the eclipse midpoint, ingress/egress times or duration.

Finally, we report mid-times in both BJD\sb{UTC} (Barycentric Julian Date, BJD, in Coordinated Universal Time) and BJD\sb{TT} (BJD\sb{TDB},  Barycentric Dynamical Time), calculated using the Jet Propulsion Laboratory (JPL) Horizons system, to facilitate handling discontinuities due to leap seconds and to allow easy comparison of eclipse mid-times (see \citealp{EastmanEtal2010apjLeapSec} for discussion of timing issues).

\newpage
\subsection{Modeling Systematics}
\label{sec:sys}

Modeling systematics is critical to recovering the extremely weak signal of an exoplanetary atmosphere against the stellar and/or background noise, particularly when using instrumentation not specifically built for the job.  Several re-analyses of early {\em Spitzer}\/ eclipse data sets underscore this.  For example, our group's initial analysis of an HD 149026b lightcurve \citet{HarringtonEtal2007natHD149026b} found two \math{\chi\sp{2}} minima, with the deeper eclipse having the deeper minimum.  This analysis used the bootstrap Monte Carlo technique as described without statistical justification and too simplistically by \citet{PressEtalNumRec}.  The re-analysis by \citet{KnutsonEtal2009apjHD149026bphase}, using MCMC, preferred the lower value, which additional observations confirmed.  Our own re-analysis, by \citealp{StevensonEtal2012apjHD149026b}, agreed with \citeauthor{KnutsonEtal2009apjHD149026bphase}.  Another example is the \citet{Desert2009} re-analysis of the putative detection of H\sb{2}O on HD 189733b by \citet{Tinetti2007Nature}.  \citeauthor{Desert2009} found a shallower transit that did not support the detection.  Although the number of such discrepancies in the {\em Spitzer}\/ eclipse and transit literature is not large compared to the many dozens of such measurements, they serve as cautionary tales.  It is critical to use only the most robust statistical treatments (e.g., MCMC rather than bootstrap), to compare dozens of systematic models using objective criteria (like BIC), and to worry about minutiae like the differences between various centering and photometry methods.  Re-analyses of photometric work done with such care have uniformly been in agreement.  Most of these appear as notes in original papers stating that another team confirmed the analysis (e.g., \citealp{StevensonEtal2012apjGJ436c}).

{\em Spitzer's}\/ IRAC channels can exhibit both time-dependent and position-dependent sensitivity variations.  These variations can be up to \sim 3\%, much more than typical (0.01\%--0.5\%) eclipse depths.  The 3.6 and 4.5 {\micron} bands use InSb detectors, and the 5.8 and 8.0 {\micron} bands use Si:As detectors.  Although each type of systematic is strongest in a different set of channels, many authors reported both systematics in both sets of channels \citep{StevensonEtal2010Natur, Reach2005-IRACCalibration, CharbonneauEtal2005apjTrES1, CampoEtal2011apjWASP12b}, so we test for them all in each observation.

The time-varying sensitivity (``ramp'') is most pronounced at 8.0 {\micron} \citep{CharbonneauEtal2005apjTrES1, HarringtonEtal2007natHD149026b} and is very weak, often nonexistent, in the InSb channels. It manifests as an apparent increase in flux with time, and at 8.0 {\microns} it is attributed to charge trapping. Observing a bright (>250 MJy\,sr\sp{-1} in channel 4), diffuse source (``preflashing'') saturates the charge traps and produces a flatter ramp \citep{KnutsonEtal2009apjHD149026bphase}.  An eclipse is easily separated from the ramp by fitting, but not without adding uncertainty to the eclipse depth. Model choice is particularly important for weak eclipses, where a poor choice can produce an incorrect eclipse depth. To model the ramp effect, we test over 15 different forms of exponential, logarithmic, and polynomial models (see \citealp{StevensonEtal2012apjHD149026b}, Equations\ (2)--(11)).

InSb detectors can have intrapixel quantum efficiency variations, which strongly affects {\em Spitzer's}\/ underresolved PSF and requires accurate (\sim 0.01-pixel) determination of the stellar center location. This intrapixel sensitivity is greatest at pixel center and declines toward the edges by up to 3.5\% \citep{Morales-Calderon2006}. It is also not symmetric about the center and the amplitude of the effect varies from pixel to pixel. Over the total duration of the observation, the position varies by several tenths of a pixel. Since the stellar center oscillates over this range frequently, this systematic is adequately sampled during a single eclipse observation. Observing with fixed pointing minimizes the effect \citep{Reach2005-IRACCalibration, CharbonneauEtal2005apjTrES1, HarringtonEtal2007natHD149026b, StevensonEtal2010Natur}.

Our BLISS method \citep{StevensonEtal2012apjHD149026b} maps a pixel's sensitivity on a fine grid of typically over 1000 ``knots'' within the range of stellar centers.  It then uses bilinear interpolation to calculate the sensitivity adjustment for each observation from the nearest knot values (\math{M(x,y)} in Equation (\ref{eqn:full})).  To compute the map, we divide the observed fluxes by the eclipse and ramp models, and assume that any residual fluxes are related to the stellar center's position in the pixel (hence the need for accurate stellar centering; see above).  We average the residuals near each knot to calculate its value.  Each data point contributes to one knot, and each knot comes from a small, discontiguous subset of the data.  The map is recalculated after each MCMC iteration and is used to calculate \math{\chi\sp{2}} in the next iteration.  The MCMC does not directly vary the knot values, but the values change slightly at each iteration.  This method quickly converges.

The crucial setup item in BLISS is determining the knot spacing (i.e., bin size or resolution). The bin size must be small enough to catch any small-scale variation, but also large enough to ensure no correlation with the eclipse fit (see Section \ref{sec:ch1}).  Either bilinear (BLI) or nearest-neighbor (NNI) interpolation can generate the sensitivities from the knots.  Assuming accurate centering, BLI should always outperform NNI. The bin size where NNI outperforms BLI thus indicates the centering precision and determines the bin size for that particular data set. If NNI always outperforms BLI, that indicates very weak intrapixel variability, and intrapixel modeling is unnecessary.

Compared to polynomial intrapixel models, the SDNR improves with BLISS mapping, but this would be expected of any model with more degrees of freedom.  Previously, we have used BIC and AIC to evaluate whether a better fit justifies more free parameters. Both BIC and AIC are approximations to the Bayes factor, which is often impractical to calculate.  Both criteria apply a penalty to \math{\chi\sp{2}} for each additional free parameter (\math{k}, in Equations\ (\ref{eqn:BIC}) and (\ref{eqn:AIC})), allowing comparison of model goodness-of-fit to the same dataset for different models.  However, both criteria assume that every data point contributes to each free parameter.  That is, they assume that changing any data point potentially changes {\em all} of the free parameters, as do all other information criteria we have researched. However, each BLISS knot value comes from only a specific, tiny fraction of the data.  Changing any individual data point changes exactly one BLISS knot.  Thus, the knots each count for much less than one free parameter in the sense of the assumptions of BIC and AIC, but not zero (i.e., they each increment \math{k} by much less than 1).  Because BLISS violates their assumptions, BIC and AIC are inappropriate for comparing models using BLISS to models that do not use it.  It is still possible to compare two models using BLISS maps with the same knot grid because the increment in the penalty terms would be the same for both grids and would thus not affect the comparison.  See Appendix A of \citet{StevensonEtal2012apjHD149026b} for a more statistically rigorous discussion.

At this point in BLISS's development, we are still working on an appropriate comparison metric. What we do know is that BLISS resolves fine detail in pixel sensitivity that, in many cases, is not compatible with any low-order polynomial form.  For example, \citet{StevensonEtal2012apjHD149026b} show (and compensate for) the effects of pixelation in digital aperture photometry, and demonstrate how our interpolated aperture photometry reduces pixelation bias.  For this paper, the eclipse-depth values are similar between BLISS and non-BLISS analyses, and the residuals are smaller with BLISS, since it is taking out some of these effects in a way that low-order polynomial models cannot (see Figure \ref{fig:SensMap} and examples in \citealp{StevensonEtal2010Natur}).  We have a large excess of degrees of freedom, so we adopt the BLISS results. We continue to use BIC for ramp-model selection.

\subsection{Modeling Light Curves and the Best Fit Criteria}
\label{sec:LightCurves}

To find the best model, for each aperture size we systematically explore every combination of ramp model and intrapixel sensitivity model.  The final light curve model is:

\begin{eqnarray}
\vspace{-10pt}
\label{eqn:full}
F(x, y, t) = F\sb{s}R(t)M(x,y)E(t),
\end{eqnarray}

\noindent where \math{F(x,y,t)} is the aperture photometry flux, \math{F\sb{\rm s}} is the constant system flux outside of the eclipse, \math{R(t)} is the time-dependent ramp model, \math{M(x, y)} is the position-dependent intrapixel model and \math{E(t)} is the eclipse model \citep{MandelAgol2002ApJtransits}.  We fit each model with a Levenberg--Marquardt \math{\chi\sp{2}} minimizer and calculate SDNR, BIC, and AIC (note that parameter uncertainties, and hence MCMC, are not needed for these calculations).

To estimate uncertainties, we use our MCMC routine with the Metropolis--Hastings random walk algorithm, running at least 10\sp{6} iterations to ensure accuracy of the result. This routine simultaneously fits eclipse parameters and {\em Spitzer} systematics.  It explores the parameter phase space, from which we determine uncertainties fully accounting for correlations between the parameters. The depth, duration, midpoint, system flux, and ramp parameters are free. Additionally, the routine can model multiple events at once, sharing the eclipse duration, midpoint and ingress/egress times. These joint fits are particularly appropriate for channels observed together (see \citealp{CampoEtal2011apjWASP12b} for more details about our MCMC routine).

To avoid fixing any model parameter during MCMC, we use Bayesian priors \citep[e.g.,][]{Gelman2002}. This is particularly relevant for noisy or low signal-to-noise (S/N) datasets where some parameters like ingress and egress times are not well constrained by the observations. For them we use informative priors taken from the literature (see Section \ref{sec:SecEclResults} for the values used in this analysis).

Photometric uncertainties used in our analyses are derived by fitting an initial model with a Levenberg--Marquardt \math{\chi\sp{2}} minimizer and re-scaling it so reduced \math{\chi\sp{2}} = 1.  This is needed because {\em Spitzer} pipeline uncertainties have often been overestimated \citep{HarringtonEtal2007natHD149026b}, sometimes by a factor of two or three. Along with the BCD frames, the Spitzer Science Center provides images giving the uncertainties of the BCD pixels.  The calculations behind these images include uncertainty in the absolute flux calibration, which effect we divide out.  The {\em Spitzer}-provided errors are thus too large for exoplanet eclipses and transits, but they do contain information about the relative noisiness of different pixels.

Most workers ignore the uncertainty frames and calculate a single per-frame uncertainty from their root mean square (rms) model residuals, sometimes taken over just a short time span.  This has the effect of fixing the reduced chi-squared to 1, and possibly ignoring red noise, depending on the time span of residuals considered.  We do use the {\em Spitzer}-provided uncertainties, resulting in slightly differing uncertainties per frame.  However, this approach can produce reduced chi-squared values of 0.3, and sometimes 0.1, as the {\em Spitzer}\/ uncertainties are computed with absolute calibration in mind.  So, we also re-scale the per-frame uncertainties to give a reduced chi-squared of 1.  As a practical matter, the variation in our uncertainties is a few percent and the typical uncertainty is the same as with the rms method applied to the entire dataset, which accounts for a global average of red noise.

Rescaling the uncertainties is changing the dataset, and BIC can only compare different models applied to a single dataset.  So, we use just one rescaling per aperture size, and fit all the models to that dataset. This works because the reasonable models for a given dataset all produce nearly the same scaling factor.  The rank ordering of models is not altered by the scaling factor. In the Section\ \ref{sec:joint} we lists the rescaling factor for each dataset.  

After deriving new uncertainties, we re-run the minimizer and then run MCMC. If MCMC finds a lower \math{\chi\sp{2}} than the minimizer, we re-run the minimizer starting from the MCMC's best value. The minimizer will find an even better \math{\chi\sp{2}}. We then restart the MCMC from the new minimizer solution.  We ensure that all parameters in four independent MCMC chains converge within 1\% according to the \citet{GelmanRubin1992} test.  We also inspect trace plots for each parameter, parameter histograms, and correlation plots for all parameter pairs.

Our measures of goodness of fit are SDNR, BIC, and AIC values \citep{Liddle2008}:  

\begin{equation}
\label{eqn:BIC}
{\rm BIC} = \chi\sp{2} + k\ln N,
\end{equation}

\begin{equation}
\label{eqn:AIC}
{\rm AIC} = \chi\sp{2} + 2k,
\end{equation}

\noindent where \math{k} is the number of free parameters, N is the number of data points. These criteria penalize additional free parameters in the system, with better fits having lower values. To appropriately compare BIC or AIC values for a given aperture size, and determine the best fit, we use the same uncertainties for each dataset, and model all combinations of ramp models and intrapixel model. SDNR values are used to compare different aperture sizes using the same model. The lowest value defines the best aperture size.

Equally important is the correlation in the residuals (see Section\ \ref{sec:ch1}). We plot and compare the scaling of binned model residuals versus bin size \citep{pont:2006, winn:2008} with the theoretical \math{1/\sqrt{N}} scaling for the rms of Gaussian residuals.  A significant deviation between those two curves indicates time-correlated variation in the residuals and possible underestimation of uncertainties if only their point-to-point variation is considered.  Note that our uncertainty estimation uses the residuals' global rms, so we already account for a global average of correlated noise.

After MCMC is finished, we study parameter histograms and pairwise correlations plots, as additional indicators of good posterior exploration and convergence.


\if\submitms y
\clearpage
\fi
\begin{figure*}[htb]
\vspace{-5pt}
\strut\hfill
\includegraphics[width=0.33\textwidth, clip]{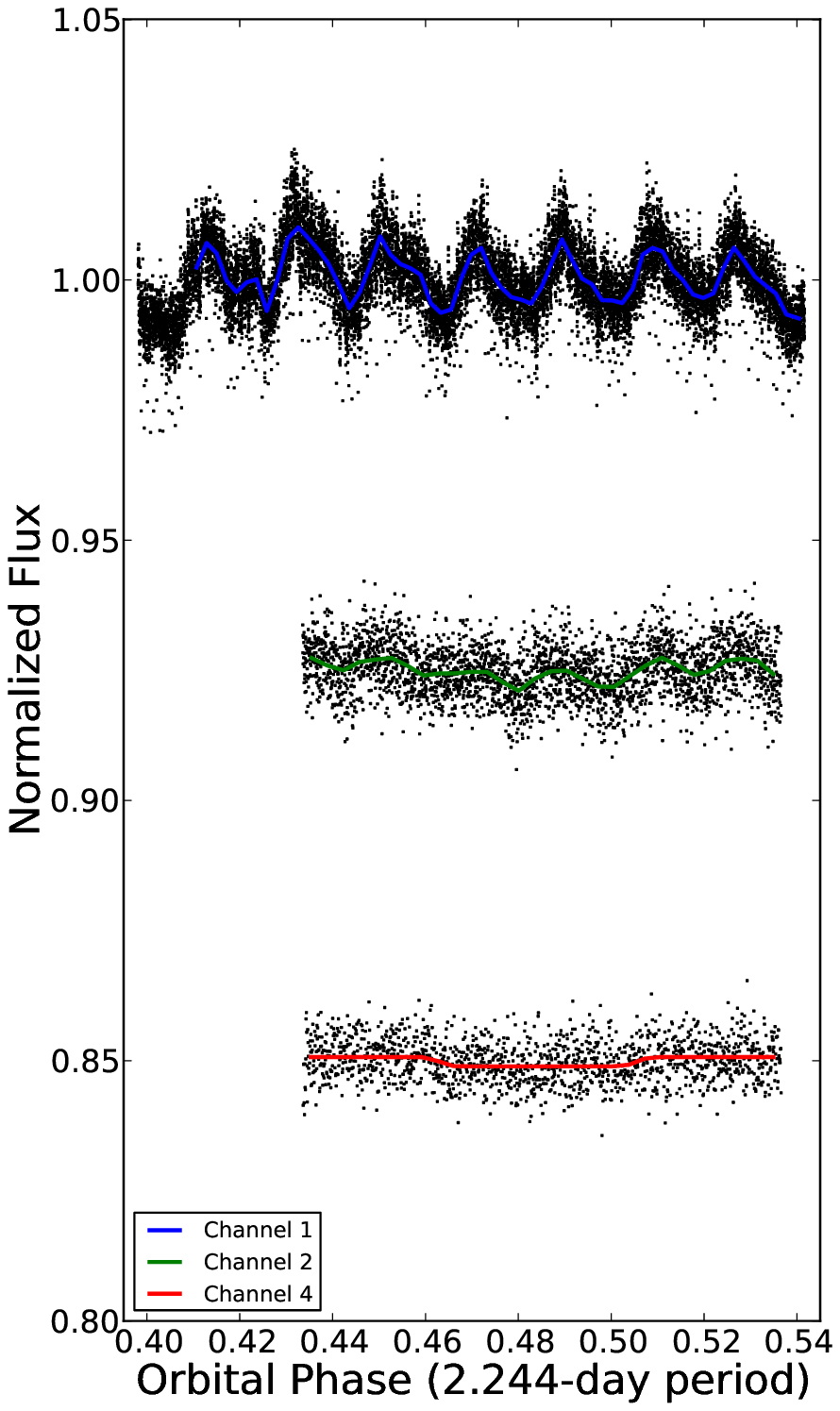}\hfill
\includegraphics[width=0.33\textwidth, clip]{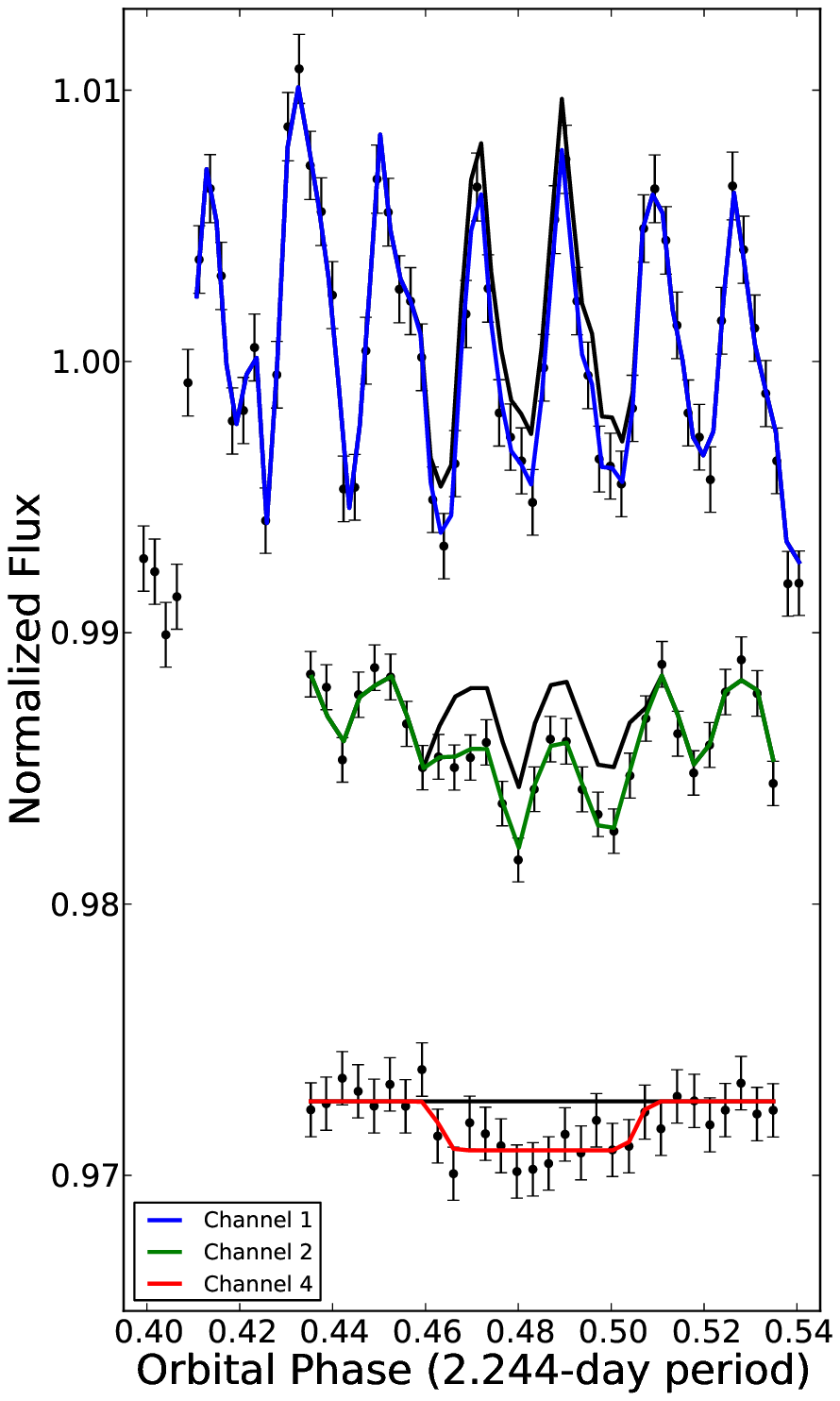}\hfill
\includegraphics[width=0.33\textwidth, clip]{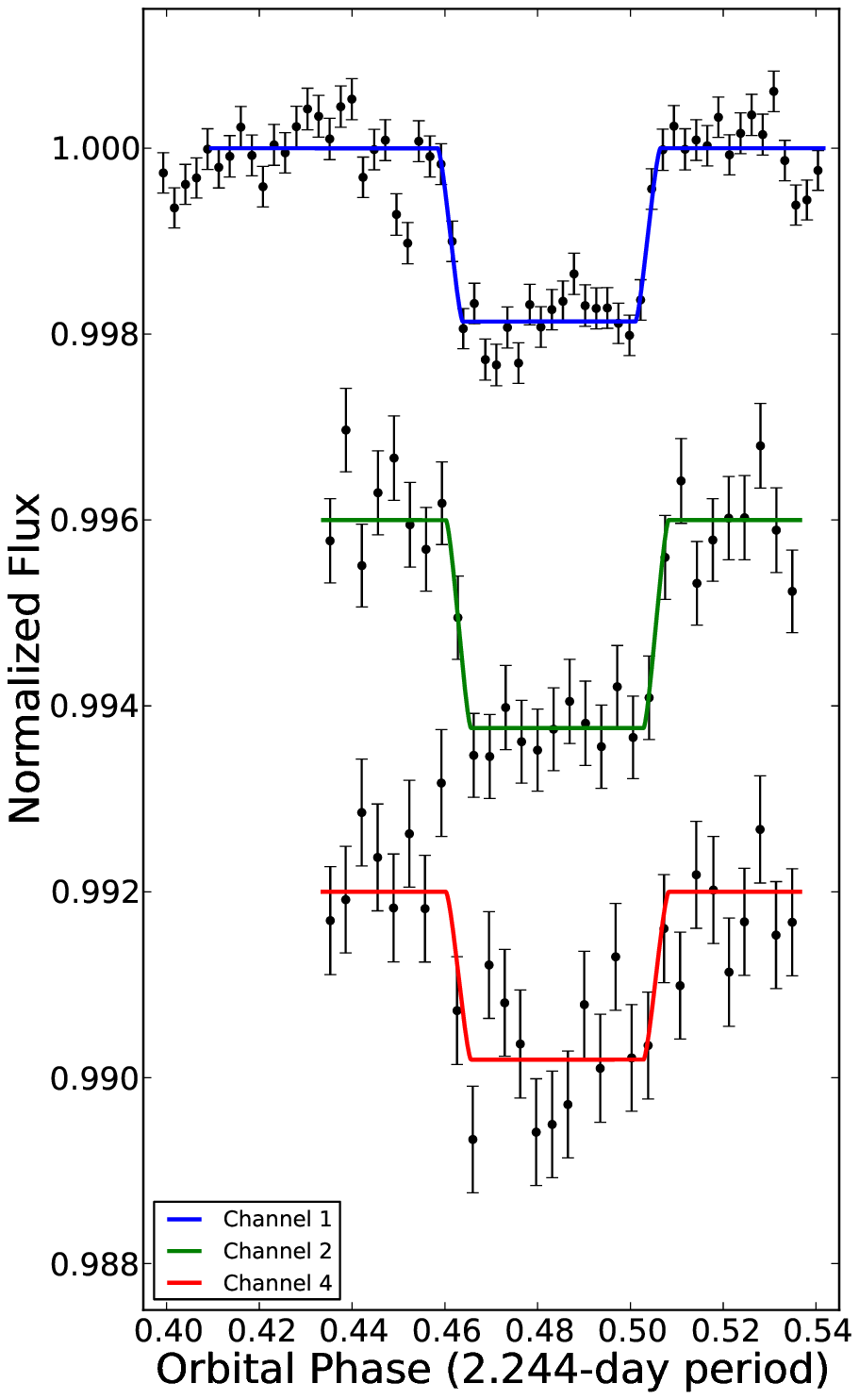}
\hfill\strut
\figcaption{\label{fig:RawBinNorm}
Raw (left), binned (center, 60 points per bin), and systematics-corrected (right) secondary-eclipse light curves of WASP-14b at 3.6, 4.5, and 8.0 {\microns}. The results are normalized to the system flux and shifted vertically for clarity.  The colored lines are best-fit models and the error bars are 1\math{\sigma} uncertainties. The black lines in the binned plots are models without an eclipse. As seen in the same plots of channels 2 and 4, a ramp model is not needed to correct for the time-dependent systematic even without clipping any initial data points. The channel 1 model omits early data due to an initial pointing drift (see Section\ \ref{sec:ch1}).
}
\end{figure*}
\if\submitms y
\clearpage
\fi

\section{WASP-14\MakeLowercase{b} PHOTOMETRY}
\label{sec:Phot}

For our analyses, we used BCD frames generated in the {\em Spitzer} IRAC pipeline \citep{Fazio2004IRAC}. The pipeline version used for each observation is given in Section\ \ref{sec:obs}. Our data reduction procedure started with applying {\em Spitzer}'s bad pixel masks and with our procedure for flagging additional bad pixels \citep{HarringtonEtal2007natHD149026b}. In each group of 64 frames and at each pixel position, we applied two-iteration outlier rejection, which calculated the frame median and the standard deviation from the median (not mean), and flagged pixels that deviated by more than 4\math{\sigma}. Then we found the stellar centroid for the photometry by using a two-dimensional Gaussian fit to data in an aperture radius of four pixels.
 
After subtracting the mean background (annuli given in Section\ \ref{sec:joint}), light curves were extracted using 5\math{\times}-interpolated aperture photometry \citep{HarringtonEtal2007natHD149026b} for every aperture radius from 2.25 to 4.25 pixels in 0.25 pixel steps.

To calculate the BJD of each exposure we used the mid-exposure time of each frame, based on the UTCS-OBS value in the FITS header and the frame number.  We performed our barycentric light-time correction using our own code and the coordinates of the {\em Spitzer} spacecraft from the Horizons ephemeris system of the JPL.  The times are corrected to BJD\sb{TDB} to remove the effects of leap seconds and light-travel time across the exoplanet's orbit.

\section{WASP-14\MakeLowercase{b} SECONDARY ECLIPSES}
\label{sec:SecEclResults}

Here, we discuss each channel's analysis and model selection in detail, particularly focusing on channel 1, due to the demanding analysis of that data set. In Subsection \ref{sec:ch1} we give our control plots, as an example of how we verify that our results are indeed the best solution for the particular data set. We present each channel separately, followed by a joint fit to all data. Figure \ref{fig:RawBinNorm} shows our best-fit eclipse light curves. Figure \ref{fig:rms} shows how the rms of the residuals scales with bin size, a test of correlated noise. In the Appendix we summarize parameters for the WASP-14 system as derived from this analysis and found in the literature.

\subsection{Channel 1--3.6 {\micron}}
\label{sec:ch1}

The channel-1 observation lasted 7.8 hr, giving ample baseline before and after the secondary eclipse.  The telescope drifted at the start of the observation.   Models with initial data points removed produce better fits with lower values for SDNR.  We therefore ignored some initial data (\sim 36 minutes, 1100 of 13760 points).  Figure \ref{fig:ch1-SDNR} compares SDNR values for models with different ramps and with and without exclusion of the initial data.

Starting from an aperture radius of 2.25 pixels and continuing in increments of 0.25 pixels, we tested all of the ramp models (linear, rising, exponential, sinusoidal, double exponential, logarithmic, etc.). Corresponding equations are listed in \citet{StevensonEtal2012apjHD149026b}. To determine the best solution we consider our best-fit criteria (see Section \ref{sec:LightCurves}) and study the correlation plots. Most of the models produced obvious bad fits, so minimizer and shorter MCMC runs eliminated them. The best aperture radius is 2.75 pixels (see Figure \ref{fig:ch1-SDNR}, bottom panel). We tested the dependence of eclipse depth on aperture radius \citep{AndersonEtal010ApJ-WASP17b}. The trend in some events may indicate a slightly imperfect background removal (see Figure \ref{fig:ch1-depths}).  The effect is less than 1\math{\sigma} on the eclipse depth.

\begin{figure*}[ht]
\vspace{-5pt}
    \centering
    \includegraphics[width=0.93\linewidth, clip]{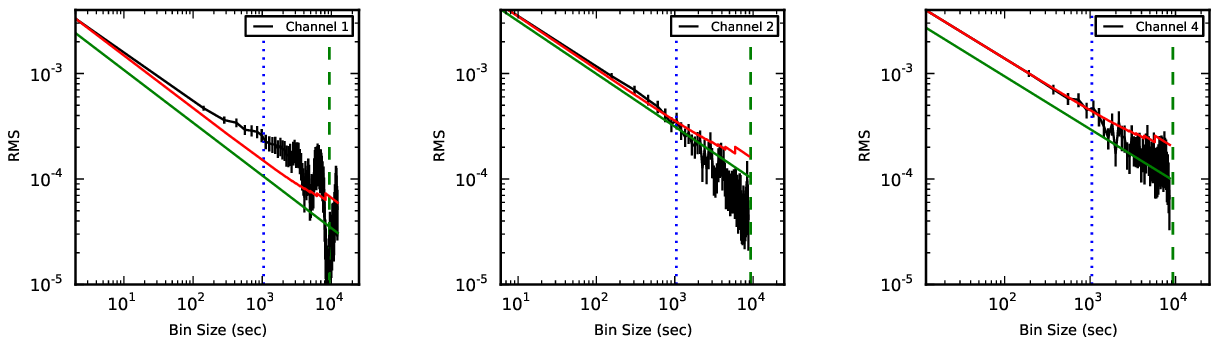}
\vspace{-10pt}
\caption{Correlations of the residuals for the three secondary eclipse light curves of WASP-14b, following \citet{pont:2006}. The black line represents the rms residual flux vs. bin size. The red line shows the predicted standard error scaling for Gaussian noise. The green line shows the Poisson-noise limit. The black vertical lines at each bin size depict 1\math{\sigma} uncertainties on the rms residuals, \math{rms/\sqrt{2N}}, where N is the number of bins (see \citealp[][Section 3.41]{Jeffreys61} and \citealp[][Section 3.3]{Sivia06} for  a derivation including the factor of two, which arises because this is the uncertainty scaling of the rms, not the mean). The dotted vertical blue line indicates the ingress/egress timescale, and the dashed vertical green line indicates the eclipse duration timescale. Large excesses of several \math{\sigma} above the red line would indicate correlated noise at that bin size. Inclusion of 1\math{\sigma} uncertainties shows no noise correlation between the ingress/egress and eclipse duration timescales anywhere except for channel 1 ingress/egress, which hints 3\math{\sigma} at a correlation (adjacent points on this plot are themselves correlated).  Since the relevant timescale for eclipse depths is the duration timescale, we do not scale the uncertainties.  See Section\ \ref{sec:WASP14act} for further discussion.
}
\label{fig:rms}
\end{figure*}

\begin{figure}[ht]
    \centering
    \vspace{-10pt}
    \includegraphics[width=0.99\linewidth, clip]{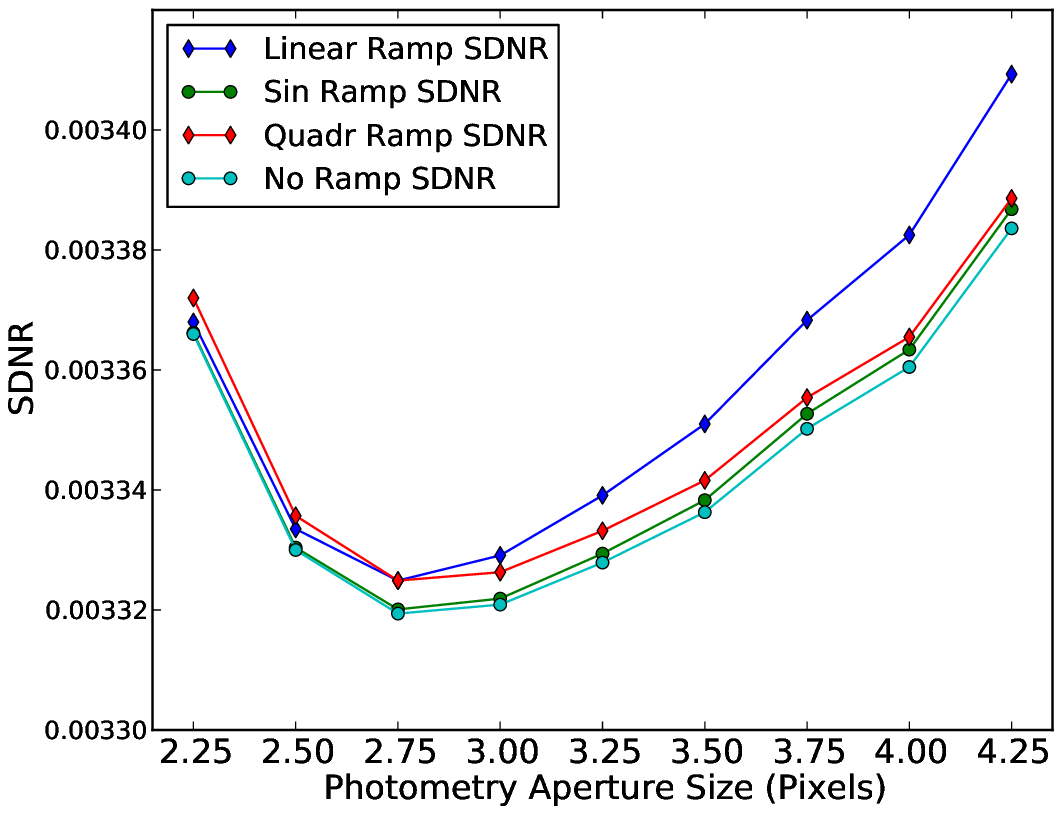}\vspace{-3pt}
    \includegraphics[width=0.99\linewidth, clip]{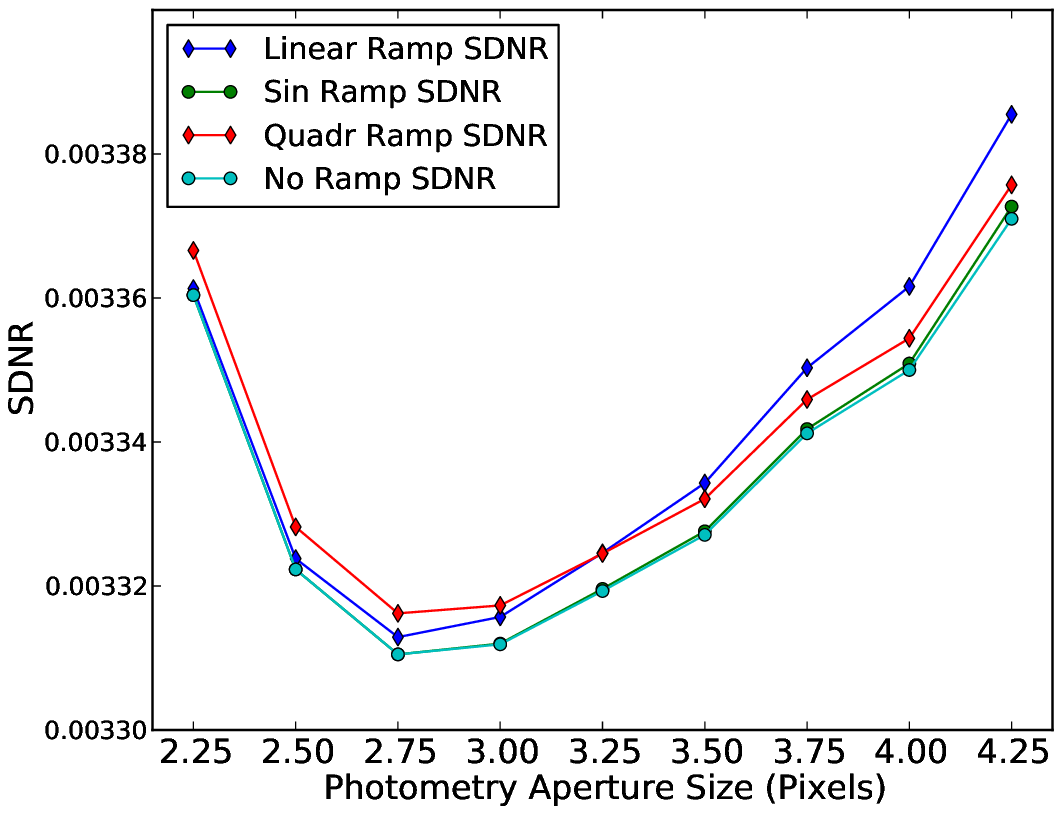}
\caption{
SDNR vs. aperture size for different ramp models in channel 1.  A lower value indicates a better model fit. Top: all observational points included ({\em no-preclip}). 
Bottom: same, but with 1100 initial points excluded ({\em preclip}).
}
\label{fig:ch1-SDNR}
\end{figure}

\begin{figure}[ht]
    \vspace{-5pt}
    \centering
    \includegraphics[width=0.99\linewidth, clip]{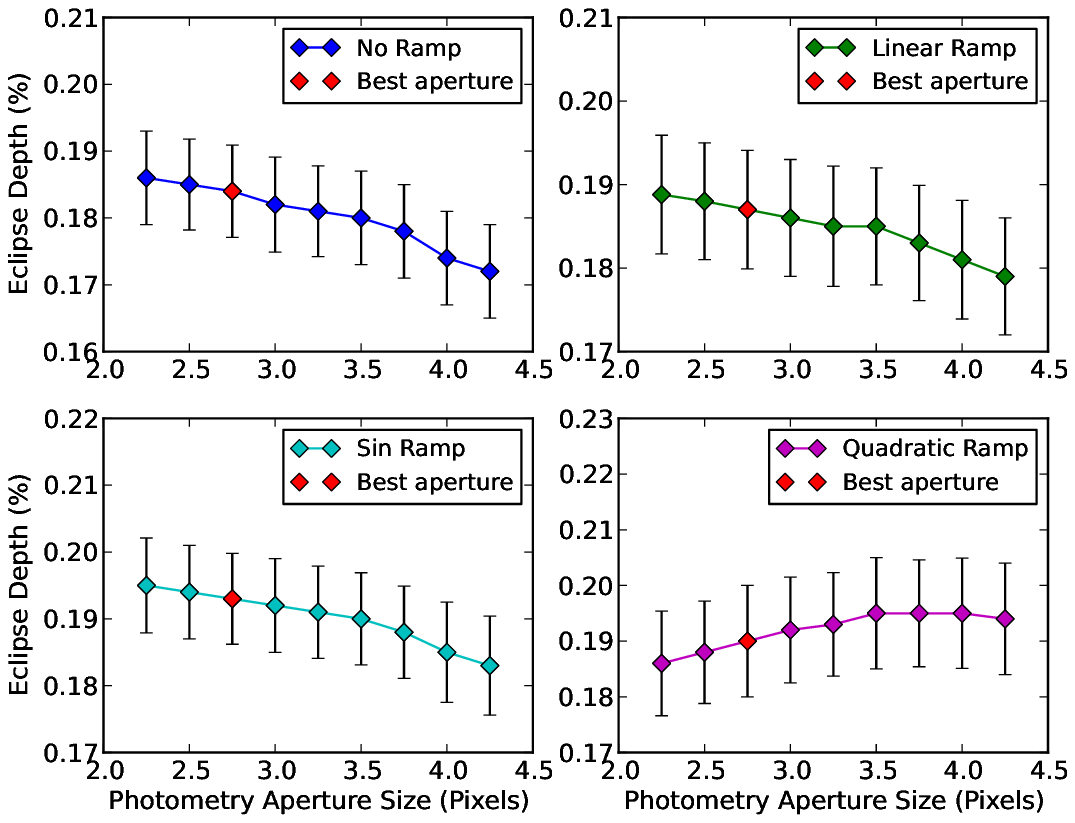}
\caption{
Best-fit eclipse depths as a function of photometry aperture size for channel 1. The four best ramp models are plotted (see bellow). The red point indicates the best aperture size for that channel. The eclipse-depth uncertainties are the result of 10\sp5 MCMC iterations. The trend shows insignificant dependence of eclipse depth on aperture size (less than 1\math{\sigma}).
}
\label{fig:ch1-depths}
\end{figure}

\begin{table}[ht]
\vspace{-10pt}
\caption{\label{table:BLISSPoly} 
Comparison BLISS And Best Polynomial Model}
\atabon\strut\hfill\begin{tabular}{lccccc}
    \hline
    \hline
    Ramp Model            & \mctc{  BLISS  }                      & \mctc{  Polynomial-Quadratic  }  \\
                  &   SDNR             &    BIC              & SDNR                   & BIC          \\
    \hline
    No ramp       & 0.003313       & 12350.0                 & 0.0033853              & 12593.2      \\ 
    Linear        & 0.003311       & 12342.3                 & 0.0033852              & 12588.5      \\ 
    Sinusoidal    & 0.003316       & 12342.2                 & 0.0033855              & 12590.5      \\
    Quadratic     & 0.003310       & 12351.5                 & 0.0033850              & 12597.3      \\
    \hline
\end{tabular}\hfill\strut\ataboff
\end{table}

\begin{figure}[ht]
    \vspace{-10pt}
    \centering
    \includegraphics[width=0.99\linewidth, clip]{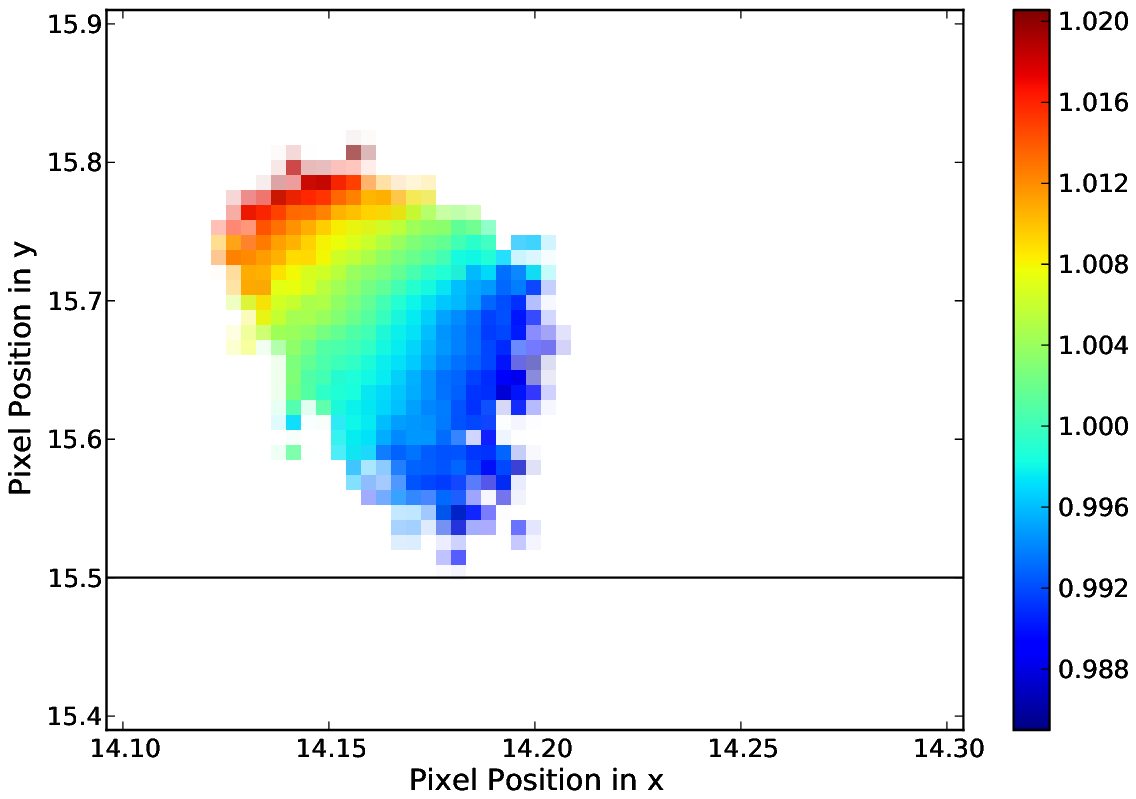}\vspace{-3pt}
    \includegraphics[width=0.99\linewidth, clip]{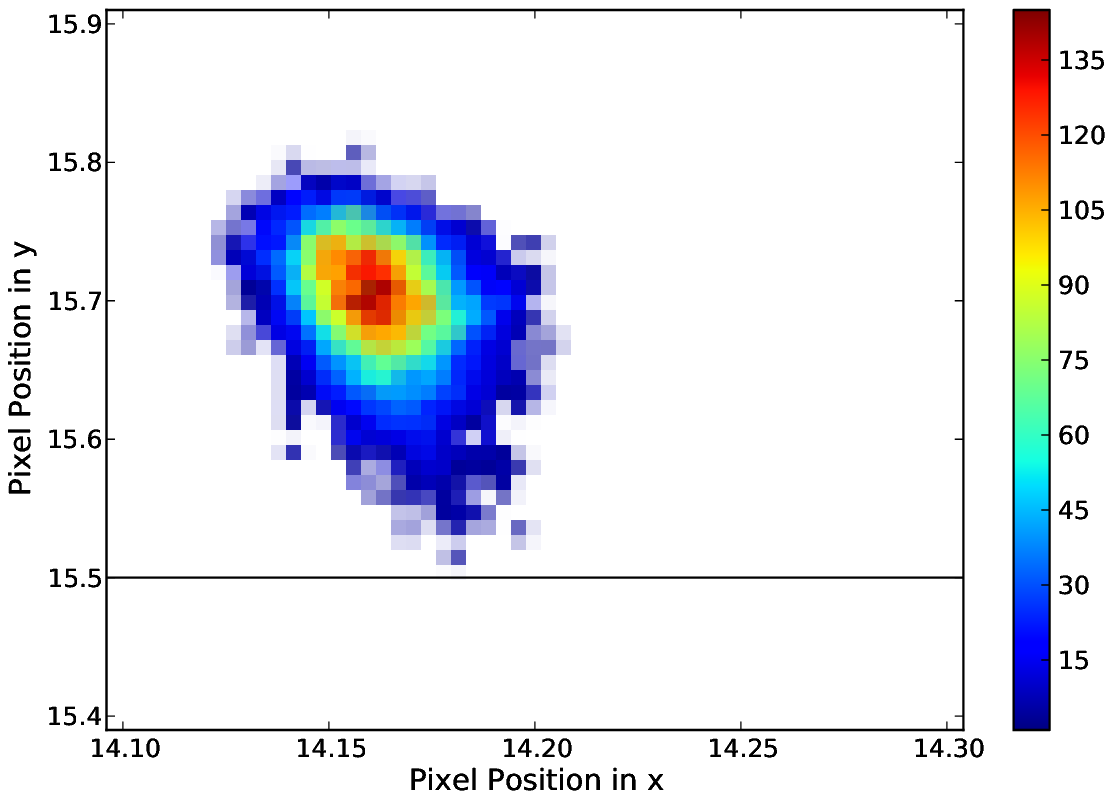}
\caption{
Top: BiLinearly Interpolated Subpixel Sensitivity (BLISS) map of channel 1.  Redder (bluer) colors indicate higher (lower) subpixel sensitivity.  The horizontal black line defines the lower pixel boundary.
Bottom: Pointing histogram.  Colors indicate the number of points in a given bin.
}
\label{fig:SensMap}
\vspace{-5pt}
\end{figure}

\begin{figure}[ht]
\vspace{-5pt}
    \centering
    \includegraphics[width=0.99\linewidth, clip]{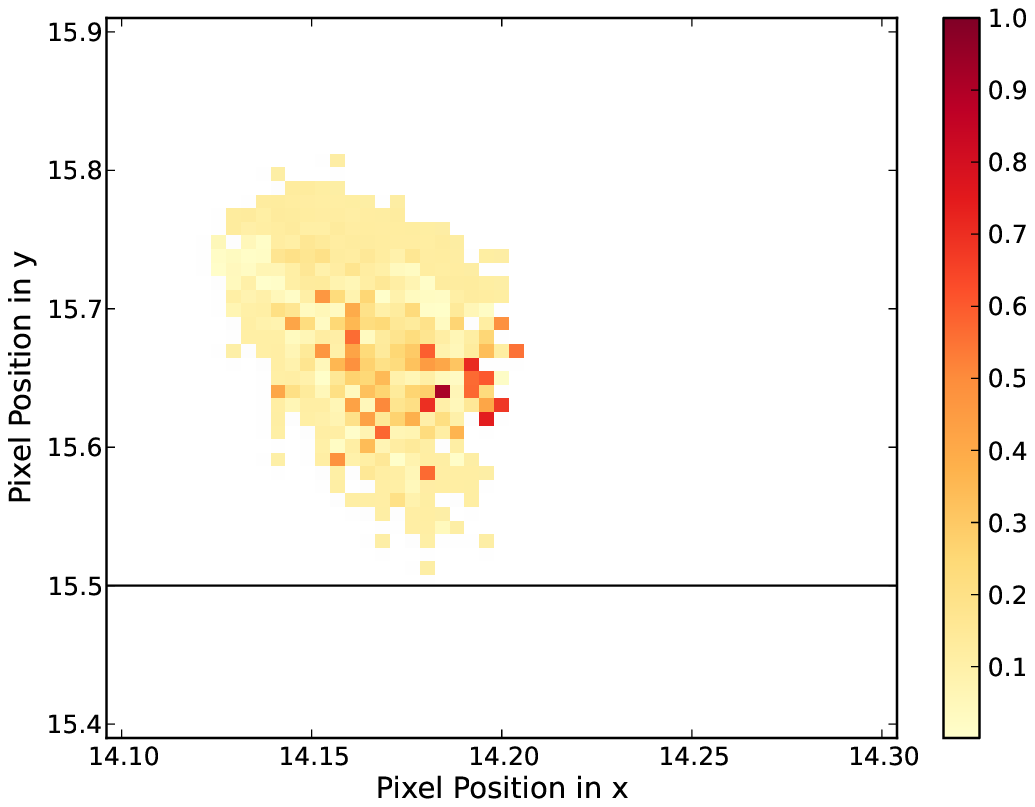}\vspace{-3pt}
\caption{
Correlation coefficients between eclipse depth and computed BLISS map knots for channel 1.  The correlation regions (in red) indicate that it is necessary to compute the BLISS map at each MCMC step, to assess the uncertainty on the eclipse depth correctly.
}
\label{fig:CorrCoef}
\end{figure}

\begin{figure}[ht]
    \centering
    \includegraphics[width=0.99\linewidth, clip]{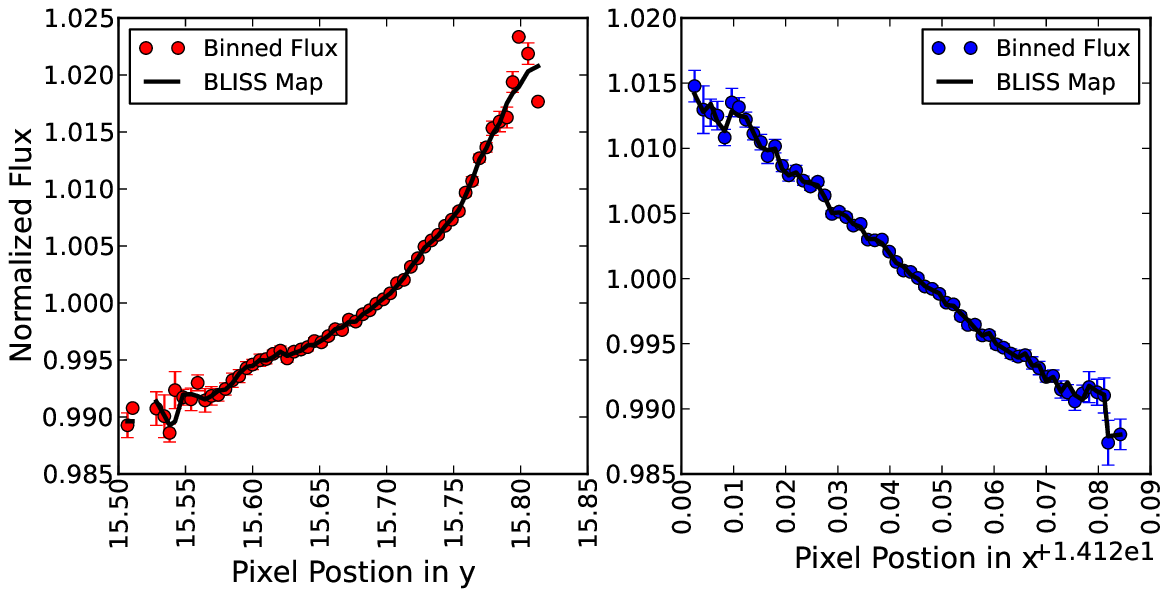}
\caption{BLISS map and data of channel 1 integrated along the $x$ (right) and $y$ (left) axes.  BLISS effectively fits the position-dependent sensitivity variation.
}
\label{fig:intrapix}
\end{figure}

Figure \ref{fig:SensMap} presents the channel-1 BLISS map and Figure \ref{fig:CorrCoef} gives the correlation coefficients between the knot values and the eclipse depth.  As stated in the Section\ \ref{sec:sys}, the most important variable to consider with BLISS is the bin size, which defines the resolution in position space. The position precision for channel 1, measured as the rms of the position difference on consecutive frames, is significantly different for the $x$ and $y$ axes (see Figure \ref{fig:intrapix}). We considered a range of bin sizes for both BLI and NNI around the calculated precision. The best bin size for this data set, determined when NNI outperformed BLI, is 0.004 pixels for $x$ and 0.01 for $y$.

We also tested two-dimensional polynomial intrapixel models \citep{Knutson08, StevensonEtal2010Natur, CampoEtal2011apjWASP12b}: 
\vspace{-10pt}
\begin{eqnarray}
\label{eq:vip} V\sb{\rm{IP}}(x,y) = p\sb1y\sp2 + p\sb2x\sp2 + p\sb3xy +
p\sb4y + p\sb5x + 1,
\end{eqnarray} 
\noindent where \math{x} and \math{y} are 
relative to the pixel center nearest the median position and
\math{p\sb{\rm 1}}--\math{p\sb{\rm 5}} are free parameters.  As noted in Section \ref{sec:sys}, we currently lack a quantitative model-selection criterion between polynomial and BLISS intrapixel models, but BIC can apply within a group of BLISS models with the same grid.  BLISS reduces SDNR significantly compared to polynomial models (see Table \ref{table:BLISSPoly}), but so would many models with more free parameters.  We use BLISS because it can handle variations that polynomials cannot follow.  See \citet{StevensonEtal2012apjHD149026b} for other tests that compare polynomial and BLISS intrapixel models.


\begin{table}[ht]
\vspace{-10pt}
\caption{\label{table:Ch1-ramps} 
Channel 1 Ramp Models}
\atabon\strut\hfill\begin{tabular}{lcccc}
    \hline
    \hline
    Ramp Model    & SDNR            & BIC            & Eclipse Depth (\%)             \\
    \hline
    No ramp       & 0.0033129       & 12350.0        & 0.184 {\pm} 0.007             \\ 
    Linear        & 0.0033105       & 12342.3        & 0.187 {\pm} 0.007             \\ 
    Sinusoidal    & 0.0033162       & 12342.2        & 0.193 {\pm} 0.007             \\
    Quadratic     & 0.0033105       & 12351.5        & 0.190 {\pm} 0.010             \\
    \hline
\end{tabular}\hfill\strut\ataboff
\end{table}

To determine the uncertainties in the model parameters, we explored the posterior probability distribution of the model given the data with MCMC. 
We used a Bayesian informative prior for the secondary-eclipse ingress and egress time (\math{t\sb{\rm{2-1}}} = 1046.8 {\pm} 43.9 s), calculated from unpublished WASP photometric and radial-velocity data. All other parameters (eclipse midpoint, eclipse duration, eclipse depth, system flux, and ramp parameters) were left free.

Considering all the above criteria (see also Section \ref{sec:LightCurves}), we selected four ramp models (see Table\ \ref{table:Ch1-ramps}). The first is without a ramp model, while the other three are: 

\vspace{-8pt}
\begin{eqnarray}
\label{eq:linramp}
R(t) = 1 + r\sb{0}\,(t - 0.5),
\end{eqnarray}

\vspace{-8pt}
\begin{eqnarray}
\label{eq:sinramp}
R(t) = 1 + a\,\rm{sin}\,(2\pi(t-t\sb{1})) + b\,\rm{cos}\,(2\pi(t-t\sb{2})),
\end{eqnarray}

\vspace{-8pt}
\begin{eqnarray}
\label{eq:quadramp}
R(t) =  1 + r\sb{1}\,(t-0.5) + r\sb{2}\,(t-0.5)\sp{2},
\end{eqnarray}
\noindent where \math{t} is orbital phase and \math{a}, \math{b}, \math{r\sb{0},r\sb{1}} and \math{r\sb{2}} are free parameters.

The models produce almost identical SDNR values. However, upon studying the BIC values and the inconsistent trend in the eclipse depths between models with similar BIC values (see Table\ \ref{table:Ch1-ramps}), we concluded that there is no single best ramp model for this data set.

Therefore, we again use Bayes's theorem and the BIC approximation to the Bayes factor to compare two different models to the data. Following \citet{Raftery1995-BIC} Equations\ (7) and (8), we calculate the posterior odds, i.e., to which extent the data support one model over the other:

\vspace{-8pt}
\begin{eqnarray}
\label{eq:odds}
\rm{Posterior\,\,\,Odds = Bayes\,\,\,Factor\,\,x\,\, Prior\,\,\,Odds},
\end{eqnarray}

\vspace{-8pt}
\begin{eqnarray}
\label{eq:prob}
\frac{P(M\sb{2}\,|\,D)}{P(M\sb{1}\,|\,D)} = \frac{P(D\,|\,M\sb{2})}{P(D\,|\,M\sb{1})}\,\,\frac{P(M\sb{2})}{P(M\sb{1})},
\end{eqnarray}

\noindent where \math{M\sb{1}} and \math{M\sb{2}} denote two models, and D denotes the data. \math{P(M\sb{1}\,|\,D)} and \math{P(M\sb{2}\,|\,D)} denote the posterior distributions of the models given the data, \math{P(D\,|\,M\sb{1})} and \math{P(D\,|\,M\sb{2})} denote the marginal probabilities of the data given the model, and \math{P(M\sb{1})} and \math{P(M\sb{2})} denote the prior probabilities of the models.

The first term on the right side of Equation\ (\ref{eq:prob}) is the {\em Bayes factor} for model 2 against model 1, which we will denote as \math{B\sb{21}}. If \math{B\sb{21}} > 1, the data favor model 2 over model 1, and vise versa. 

\citet[see his Equations (20)--(22)]{Raftery1995-BIC} further derives an approximation to the Bayes factor, using BIC, that defines the ratio of marginal probabilities for the two models as:

\vspace{-8pt}
\begin{eqnarray}
\label{eq:deltaBIC}
B\sb{21} = \frac{P(D\,|\,M\sb{2})}{P(D\,|\,M\sb{1})} \approx e^{-{\Delta {\rm BIC}}/{2}},
\end{eqnarray}

\noindent where \math{\Delta}BIC = BIC\math{(M\sb{2})} - BIC\math{(M\sb{1})}.
We calculate this quantity for each of our ramp models.


\begin{table}[ht]
\vspace{-10pt}
\caption{\label{table:deltaBIC} 
Bayes Factor for Model 2 against Model 1}
\atabon\strut\hfill\begin{tabular}{lcccc}
    \hline
    \hline
    Ramp Model    & BIC            & \math{\Delta} BIC     &   \math{B}\sb{21}        & 1\,/\,\math{B}\sb{21}      \\
    \hline
    No ramp       & 12350.0        &     7.8               &     0.02          &  49.4           \\ 
    Linear        & 12342.3        &     0.1               &     0.95          &  1.05           \\ 
    Sinusoidal    & 12342.2        &     0.0               &     ...           &  ...             \\ 
    Quadratic     & 12351.5        &     9.3               &     0.009         &  104.6          \\
    \hline
\end{tabular}\hfill\strut\ataboff
\end{table}

Table\ \ref{table:deltaBIC} gives the probability ratio, or the Bayes factor, for each of our ramp models compared to the model with the smallest BIC value (the sinusoidal model, see Table\ \ref{table:Ch1-ramps}).  These models are all within the 3\math{\sigma} confidence interval of the best model, indicating an ambiguous situation.  In the atmospheric modeling below, we use the eclipse depth and uncertainty from each of the two extreme models (no-ramp and sinusoidal), and show that the resulting atmospheric models are consistent with each other.  A representative single eclipse depth and uncertainty that spans the two points from the joint fit model (see Section \ref{sec:joint}) is 0.19\% {\pm} 0.01\%, and the corresponding brightness temperature is 2242 {\pm} 55 K.

\subsubsection{On WASP-14 Activity}
\label{sec:WASP14act}

\begin{figure}[ht]
    \vspace{-10pt}
    \centering
    \includegraphics[width=0.99\linewidth, clip]{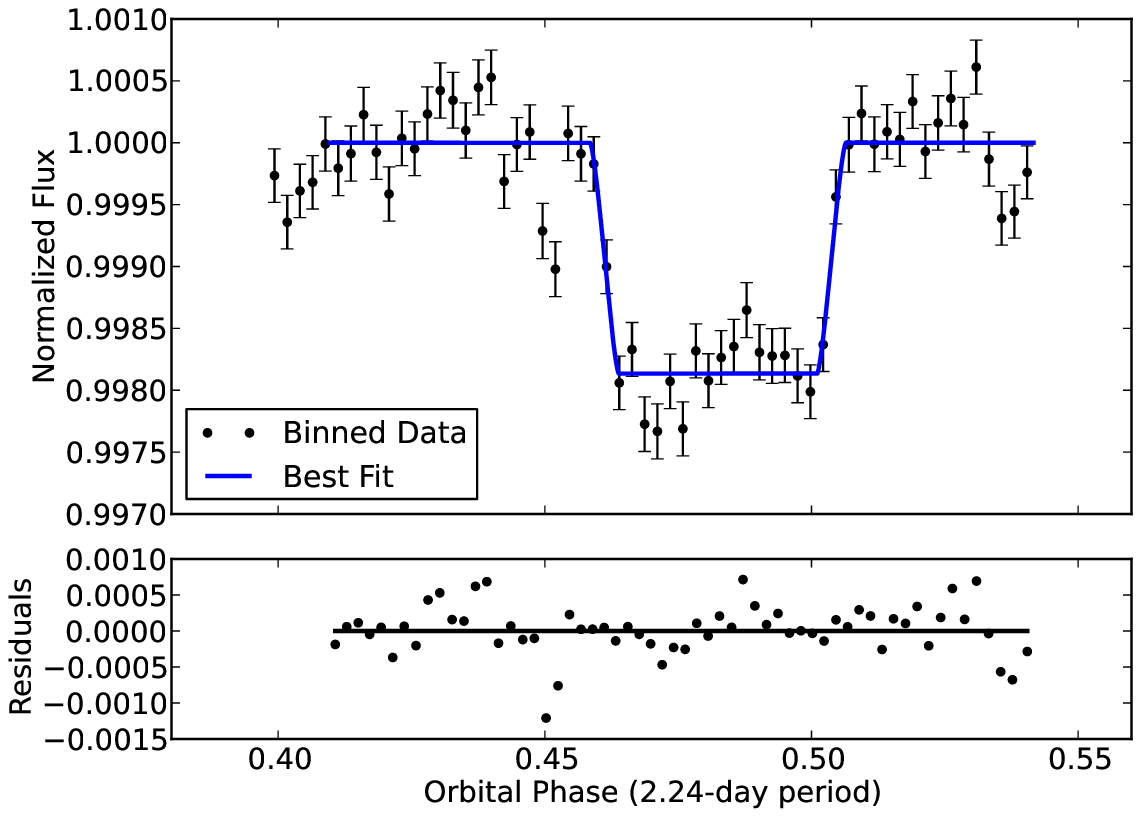}
\caption{Residuals for the channel 1 observations (lower panel) display some level of correlated noise both in and out of the eclipse.
}
\label{fig:resid}
\end{figure}

In this channel, we detect time correlation of noise at the 3\math{\sigma} level on time scales of \math{<\ttt{3}} s and \math{\lesssim2\sigma} up to about the 3000 s scale (Figure\ \ref{fig:rms}, left panel, and Figure\ \ref{fig:resid}). The longest time scale with even a \math{2\sigma} detection of correlation is about 1/7 the eclipse duration, so we do not expect a major effect on the planetary results.  Although not perfect, our ramp and intrapixel models typically remove instrumental effects (e.g., see the middle and right panels of Figure \ref{fig:rms}), raising the question of stellar activity.

One would not expect a hot mid-F star (with a small convective zone) like WASP-14 to be active or to show much spot activity even if it were a moderate rotator.  Nonetheless, we analyzed the WASP light curve of WASP-14 to determine whether it shows periodic modulation due to the combination of magnetic activity and stellar rotation.  The stellar rotation values derived by \citet{Joshi2009-WASP14b} together with the estimated stellar radius imply a rotation period of about 12 days or more, assuming that the rotation axis of the star is approximately aligned with the orbital axis of the planet. We used the sine-wave fitting method described by \citet{Maxted2011-WASP41} to calculate a periodogram over 4096 uniformly spaced frequencies from 0 to 1.5 cycles day\sp{-1}. The false-alarm probability (FAP) for the strongest peak in these periodograms was calculated using a bootstrap Monte Carlo method also described by \citet{Maxted2011-WASP41}.

We did not find any significant periodic signals (FAP < 0.05) in the WASP data, apart from frequencies near 1 cycle day\sp{-1}, which are due to instrumental effects. We examined the distribution of amplitudes for the most significant frequency in each Monte Carlo trial and used these results to estimate a 95\% upper confidence limit of 1 milli-magnitude (0.1\%) for the amplitude of any periodic signal in the lightcurve.

In our work on dozens of {\em Spitzer} eclipses, we have often found the same
channel to behave differently at different times, even on the
same star.  Our systematics removal algorithms correct the worst
effects, which are consistent, but there is sometimes still some
significant baseline scatter or oscillation.  While one might expect
certain kinds of stars to be relatively stable, {\em Spitzer} can
reach \math{\sigma\sim 0.01\%} eclipse-depth sensitivity, and
non-periodic stellar oscillations of this scale and at these wavelengths are not
well studied.  So, it is not fully clear whether these effects come
from the observatory or the star.

Since scatter and oscillation often
persist during an eclipse (when the planet is behind the star), and since a change in planetary signal of
the magnitude seen would generally mean an implausibly dramatic
change in the planet, we feel justified in treating the scatter or
oscillation phenomenologically.  In this case, our per-point uncertainties account for a global average of correlated noise.  MCMC accounts for any correlation between eclipse and model
parameters, and the rms versus bin size analysis, now including error bars, determined that the 
time correlation was not significant near the time scale of interest (Figure \ref{fig:rms}).  Also, a larger uncertainty was assigned to the eclipse depth based on model ambiguity (above), which provides an additional margin of safety.

\subsection{Channel 2--4.5 {\micron}}
\label{sec:ch2}

Channel 2 and 4 were observed at the same time. We first modeled each channel separately, determining the best aperture size, time-variability (ramp) model, and bin size for BLISS. Then we applied a joint fit. For both channels 2 and 4, we again used the Bayesian informative prior for the values of ingress and egress times (\math{t\sb{\rm{2-1}}} = 1046.8 {\pm} 43.9 s), calculated from unpublished WASP photometric and radial-velocity data. All other parameters were left free.

\begin{figure}[ht]
    \vspace{-10pt}
    \centering
    \includegraphics[width=0.99\linewidth, clip]{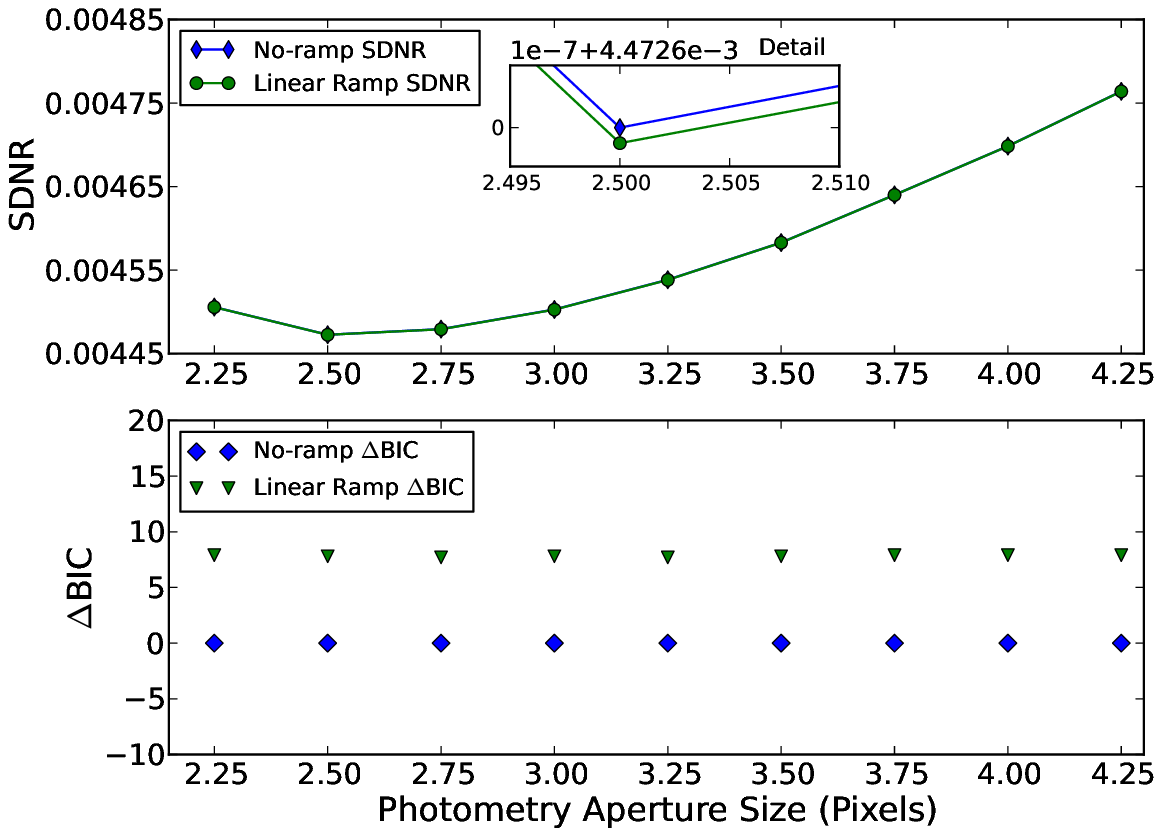}
\caption{
Channel 2 comparison between linear and no ramp models. The plots show the SDNR and \math{\Delta}BIC vs. aperture size.  A lower value indicates a better model fit.
}
\label{fig:ch2-SDNR}
\end{figure}

The observation in channel 2 lasted 5.5 hr. There was no stabilization period observed in the data, so no initial points were removed from the analysis. 

\begin{figure}[ht]
    \vspace{-10pt}
    \centering
    \includegraphics[width=0.99\linewidth, clip]{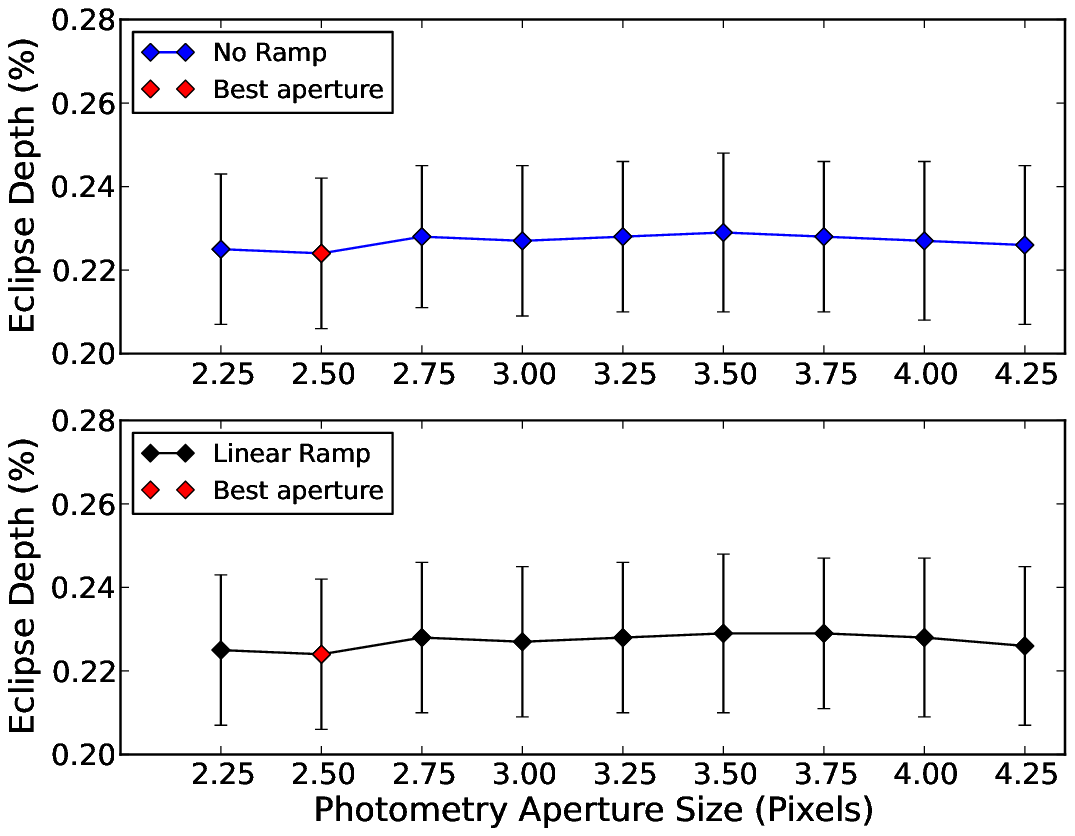}
\caption{
Best-fit eclipse depths as a function of photometry aperture size for channel 2. The red point indicates the best aperture size for that channel. The eclipse-depth uncertainties are the result of 10\sp5 MCMC iterations. The trend shows insignificant dependence of eclipse depth on aperture size (much less than 1\math{\sigma}).
}
\label{fig:ch2-depths}
\vspace{-5pt}
\end{figure}

Following the criteria in Section\ \ref{sec:backgr}, we tested each of our ramp models (Table\ \ref{table:Ch2-ramps}) at each of the aperture radii from 2.25--4.25 pixels in 0.25 pixel increments.  Figure \ref{fig:ch2-SDNR} shows SDNR and \math{\Delta}BIC versus aperture size for our two best ramp models.  We note insignificantly different SDNR values between the two ramp models, which suggests that the best dataset (aperture radius of 2.50) does not depend on the model being fit.  The BIC favors the no-ramp model. The no-ramp model is 47 times more probable than the linear model.

We also tested the dependence of eclipse depth on aperture radius (see Figure \ref{fig:ch2-depths}). The eclipse depths are well within 1\math{\sigma}.

Prior to the science observations in channels 2 and 4, we observed a 212-frame preflash (see Section\ \ref{sec:obs}) on a diffuse, uniformly bright H\sb{II} emission region centered at $\alpha~=~10,45,02.2$, $\delta~=~-59,41,10.1$. The portion of the array within the aperture of the science observation in each channel was uniformly illuminated. For channel 2, the average flux within the 2.5 pixel aperture is \sim 200 MJy\,sr\sp{-1}, while for the 3.5 pixel aperture of channel 4 it is \sim 1800 MJy\,sr\sp{-1}. 

\begin{table}[ht]
\vspace{-10pt}
\caption{\label{table:Ch2-ramps} 
Channel 2 Ramp Models}
\atabon\strut\hfill\begin{tabular}{lcccc}
    \hline
    \hline
    Ramp Model    & SDNR           & BIC           & Eclipse Depth (\%)           \\
    \hline
    No Ramp      & 0.0044726       & 2964.2        & 0.224 {\pm} 0.012           \\
    Linear       & 0.0044725       & 2971.9        & 0.224 {\pm} 0.018           \\ 
    Quadratic    & 0.0044723       & 2979.9        & 0.241 {\pm} 0.025           \\ 
    Rising       & 0.0044726       & 2980.1        & 0.224 {\pm} 0.021           \\
    Lin+Log      & 0.0044690       & 2983.9        & 0.228 {\pm} 0.017           \\
    \hline
\end{tabular}\hfill\strut\ataboff
\end{table}

\begin{figure*}[htb]
\vspace{-5pt}
\strut\hfill
\includegraphics[width=0.43\textwidth, clip]{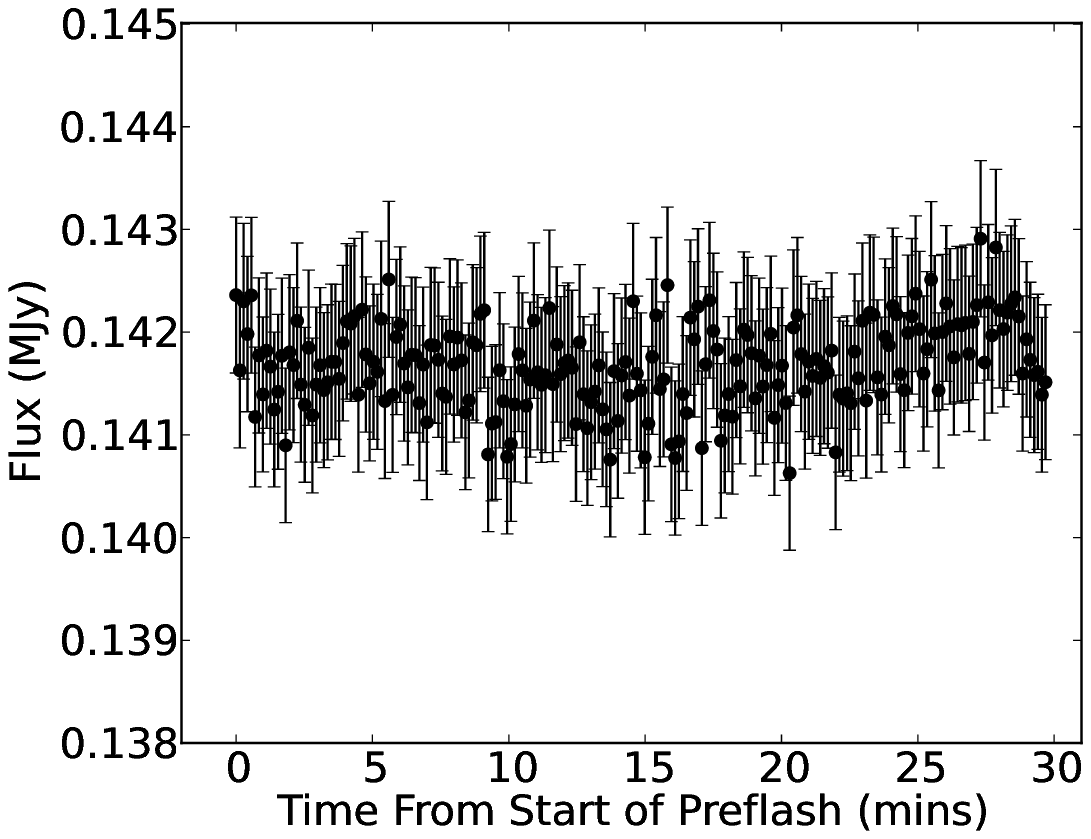}
\includegraphics[width=0.43\textwidth, clip]{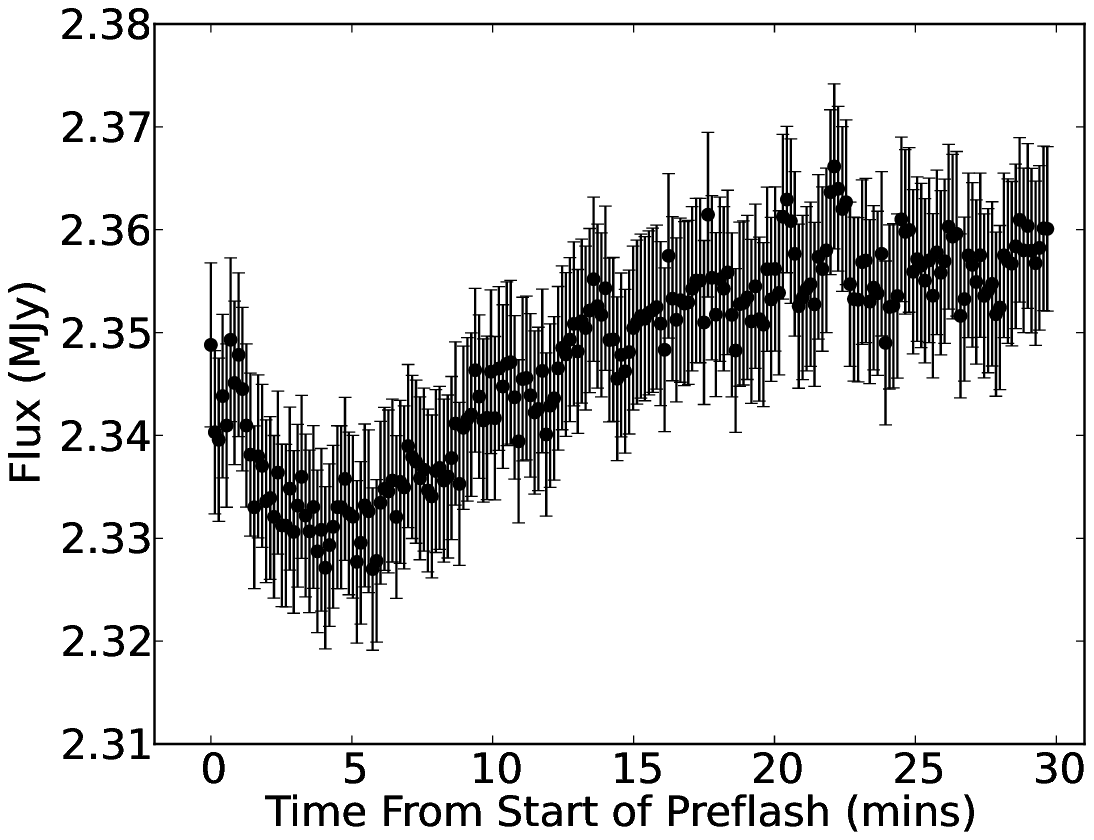}
\figcaption{\label{fig:preflash-ch24}
Preflash light curves for channel 2 (left) and channel 4 (right). The plots show binned data over 30 minutes of observation. The preflash source is a bright H\sb{II} emission region.  Without a preflash, the science observations would show a similar or possibly longer ramp in channel 4.}
\end{figure*}

As expected, channel 2 shows no increase in flux during the preflash observation (see Figure\ \ref{fig:preflash-ch24}, left panel) nor during the main science observation (see Figure \ref{fig:RawBinNorm}, raw data). The preflash observation in channel 4 saturated within the 30 minutes, eliminating the ramp effect in channel 4. 

Regardless of the preflash observations, we tested the full set of ramp equations and discarded obvious bad fits after shorter runs. Among acceptable fits, the lowest BIC value (see Table\ \ref{table:Ch2-ramps}) determined that there is no significant ramp effect in the channel 2 dataset. 

\begin{table}[h]
\vspace{-10pt}
\caption{\label{table:Ch4-ramps} 
Channel 4 Ramp Models}
\atabon\strut\hfill\begin{tabular}{lcccc}
    \hline
    \hline
    Ramp Model    & SDNR           & BIC           & Eclipse Depth (\%)           \\
    \hline
    No ramp      & 0.0039799       & 1459.2        & 0.181 {\pm} 0.013           \\
    Linear       & 0.0039770       & 1464.3        & 0.182 {\pm} 0.012           \\
    Rising       & 0.0039799       & 1466.4        & 0.198 {\pm} 0.030           \\ 
    Quadratic    & 0.0039763       & 1471.3        & 0.181 {\pm} 0.018           \\ 
    Lin+Log      & 0.0039799       & 1481.0        & 0.181 {\pm} 0.024           \\
    \hline
\end{tabular}\hfill\strut\ataboff
\vspace{-3pt}
\end{table}

\begin{figure}[ht]
    \centering
    \includegraphics[width=0.99\linewidth, clip]{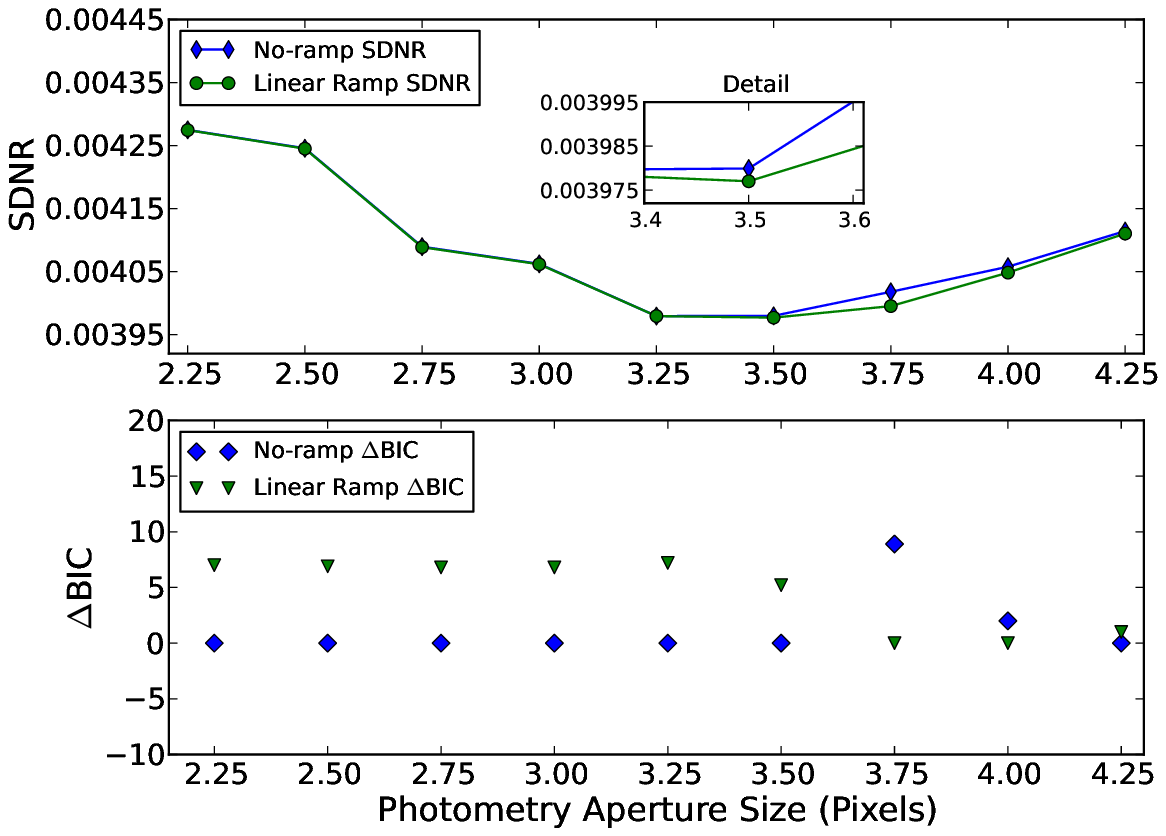}
\caption{
Channel 4 comparison between linear and no-ramp models. The plots show SDNR and \math{\Delta}BIC vs. aperture size.  A lower value indicates a better model fit.
}
\label{fig:ch4-SDNR}
\vspace{-5pt}
\end{figure}

\begin{figure}[ht]
    \centering
    \includegraphics[width=0.99\linewidth, clip]{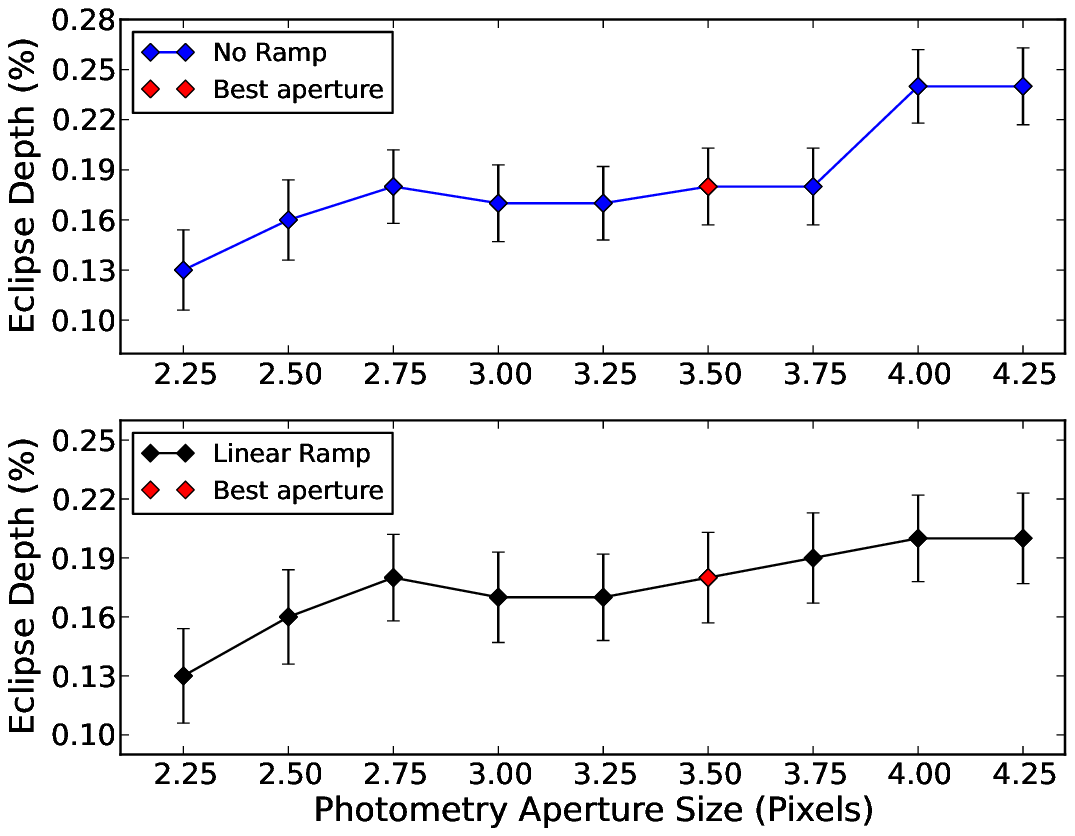}
\caption{
Best-fit eclipse depths as a function of photometry aperture size for channel 4. The red point indicates the best aperture size for that channel. The eclipse-depth uncertainties are the result of 10\sp5 MCMC iterations. This channel has the lowest S/N (\sim 8). The aperture size of 2.25 pixels shows excess noise. Excluding it, the trend exhibits insignificant dependence of eclipse depth on aperture size (less than 1\math{\sigma}).
}
\label{fig:ch4-depths}
\end{figure}

Each observation ended with a 10-frame, post-eclipse observation of blank sky in the same array position as the science observations to check for warm pixels in the photometric aperture. There were none.

To remove intrapixel variability we again apply our new BLISS technique, and also Equation\ (\ref{eq:vip}).  As with channel 1, the projection plot shows BLISS following significant variations that the polynomial does not fit well. The position precisions in channel 2 are 0.02 pixels for $x$ and 0.014 pixels for $y$. The best bin sizes are 0.028 pixels in $x$ and 0.023 pixels in $y$. The best aperture size, ramp model, and BLISS bin sizes are then used in our joint fit, which gave us the eclipse depths and the brightness temperatures in Section \ref{sec:joint}.

\subsection{Channel 4--8.0 {\micron}}
\label{sec:ch4}

Again, no stabilization period was observed in the 8.0 {\micron} dataset data set, hence no initial data points were removed. The preflash eliminated the ramp entirely, according to BIC (Table\ \ref{table:Ch4-ramps}).

Figure \ref{fig:ch4-SDNR} plots the SDNR and \math{\Delta}BIC values versus aperture size at 8.0 {\micron}. For our two best ramp models (Table\ \ref{table:Ch4-ramps}) the smallest SDNR value is at 3.50 pixels (which determined our best aperture size), and the lowest BIC value at that aperture size is for the model without a ramp. We again test for the dependence of eclipse depth on aperture size (Figure \ref{fig:ch4-depths}).


\begin{figure*}[ht]
\vspace{-5pt}
\centerline{
\includegraphics[height=5.9cm, width=5cm ]{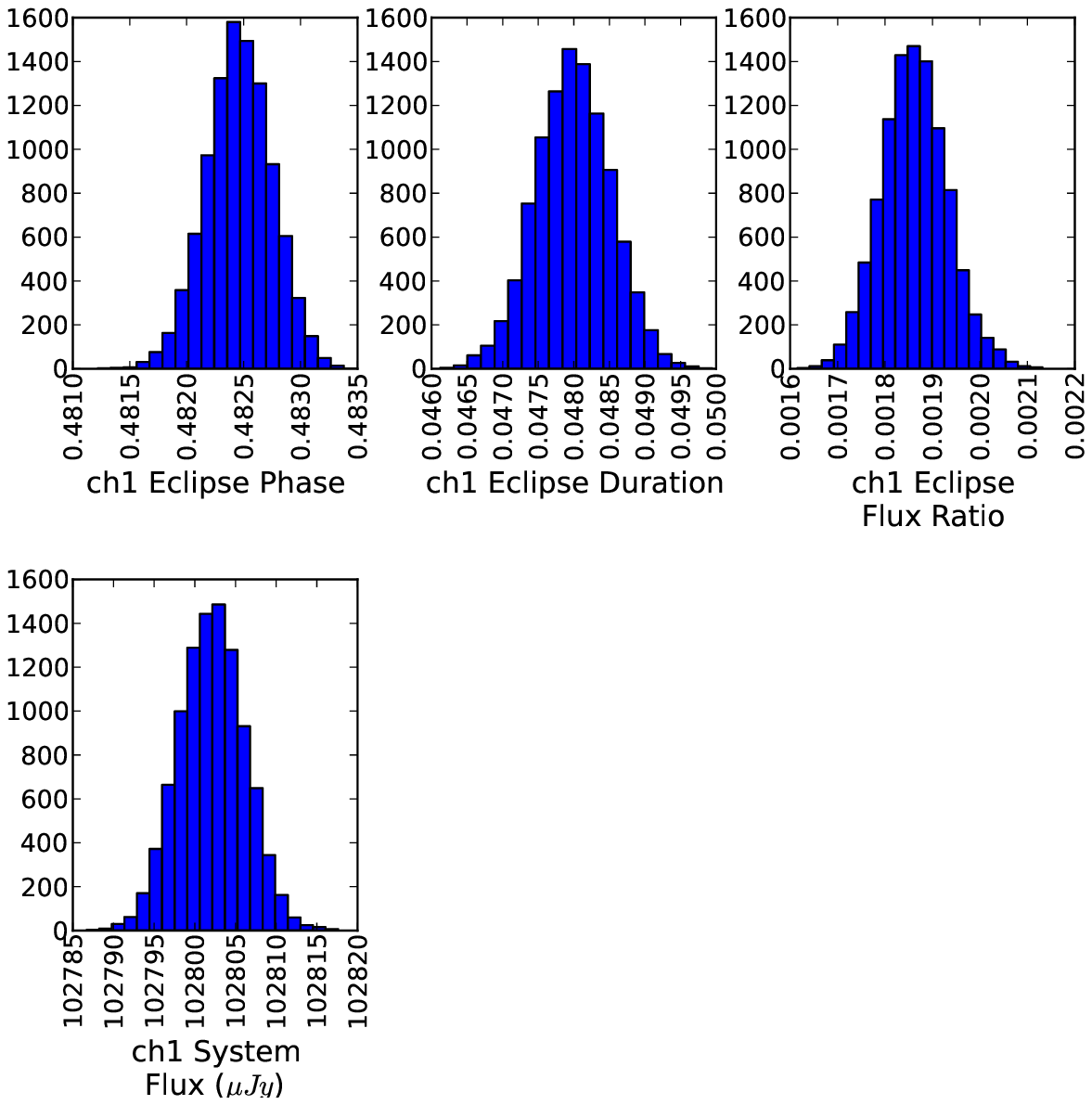}
\includegraphics[height=6cm, clip]{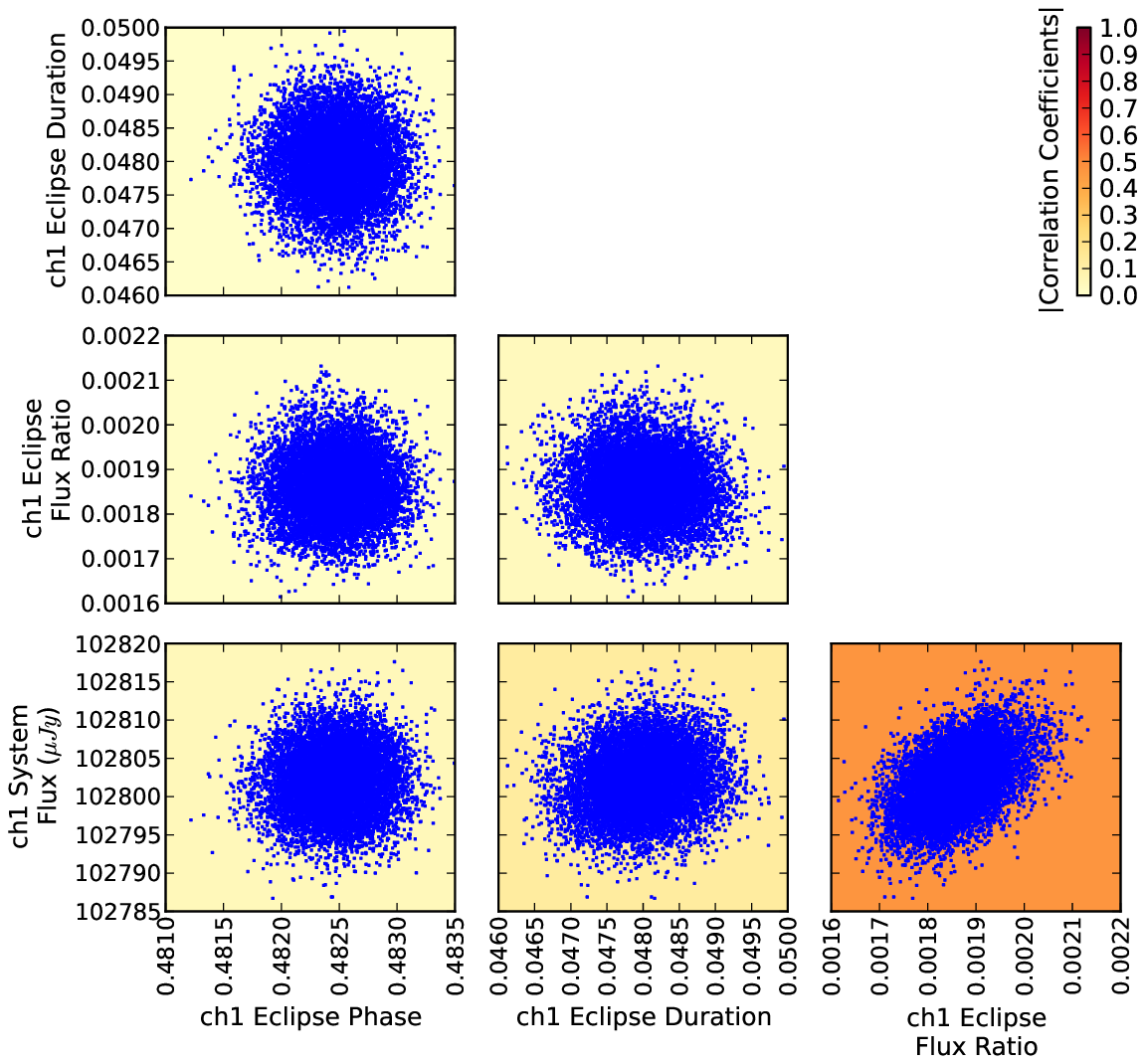}
\includegraphics[height=6cm, width=5cm ]{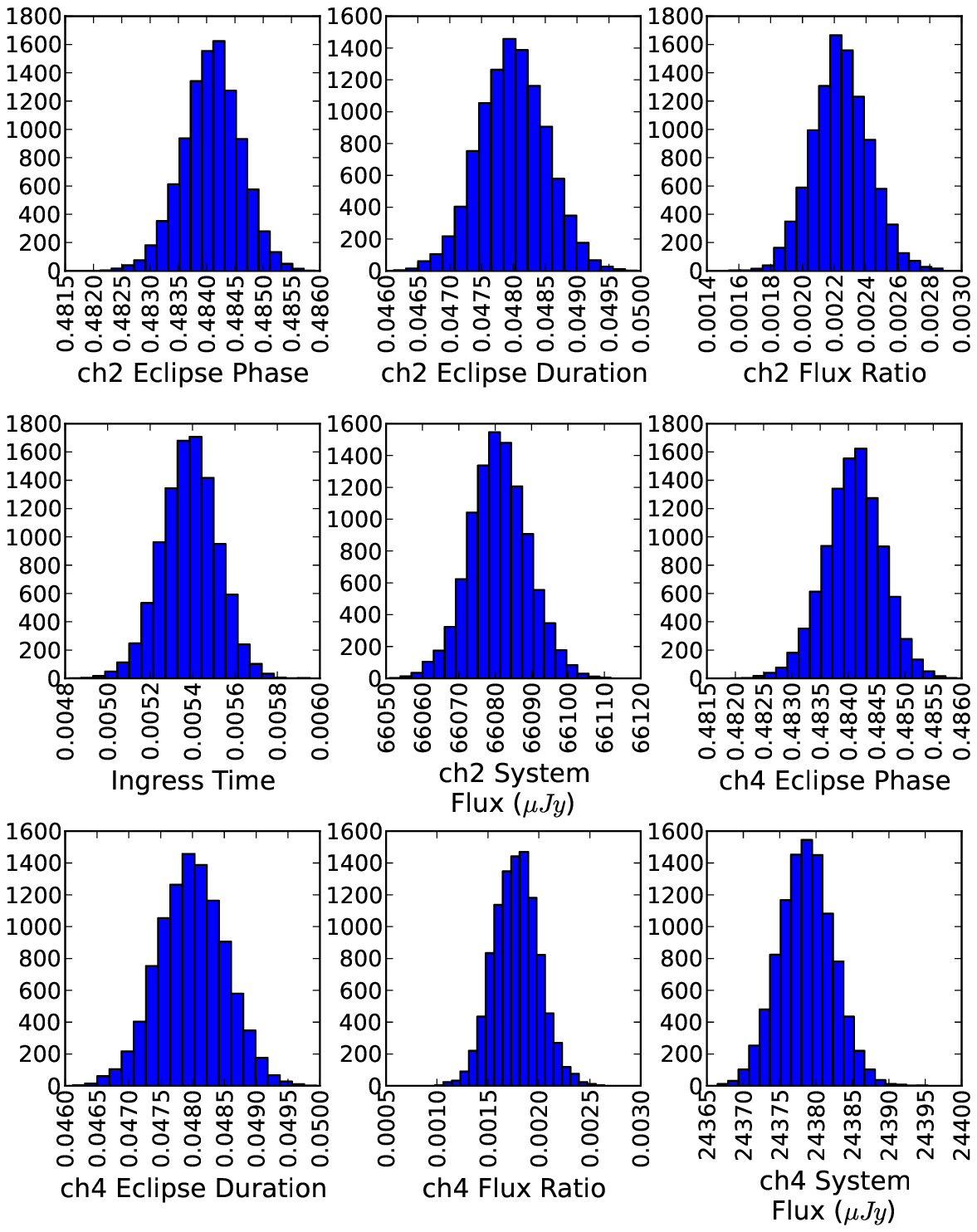}}
\caption{
Left and Center: sample parameter histograms and parameter correlations for channel 1. The background color depicts the absolute value of the correlation coefficient.
Right: sample parameter histograms for channel 2 and channel 4, produced in the joint fit. All other parameter histograms are similarly Gaussian. Every 10th step in the MCMC chain is used to decorrelate consecutive values.
}
\label{fig:Corr-Hist}
\vspace{-5pt}
\end{figure*}

Even though intrapixel variability is not so strong in channels 3 and 4, pixelation can be significant at any wavelength if the aperture is small (see \citealp{StevensonEtal2012apjHD149026b} and \citealp{Anderson2011-Ch24-WASP17b}). This justifies testing whether BLISS can give a better fit. Upon testing a full set of bin sizes, we concluded that NNI always outperforms BLI, indicating that variability from pixelation is insignificant.

\subsection{Joint Fit}
\label{sec:joint}

Our final models fit all data simultaneously.  The models shared a common eclipse duration for channels 1, 2, and 4 and a common midpoint time for channels 2 and 4, which were observed together.  We used the same priors as above.  The \citet{GelmanRubin1992} convergence diagnostic dropped below 1\% for all free parameters after 50,000 iterations.  Histograms for some interesting parameters for channel 1 appear on the left side of Figure \ref{fig:Corr-Hist}.  The middle plots show the pairwise correlations (marginal distributions) of these parameters.  The histograms on the right are for the joint fit of channels 2 and 4. All other histograms are similarly Gaussian, confirming that the phase space minimum is global and defining the parameter uncertainties.  Tables\ \ref{tab:eclfits-lin} and \ref{tab:eclfits-sin} report two joint-fit results for our two best ramp models in channel 1 (linear and sinusoidal), along with photometric results and modeling choices from the individual fits. Light-curve files including the best-fit models, centering data, photometry, etc., are included as electronic supplements to this article.


\begin{table*}[ht]
\vspace{-5pt}
\centering
\caption{\label{tab:eclfits-lin} Joint Best-fit Eclipse Light-curve Parameters (Channel 1--Linear Ramp)}
\begin{tabular}{lr@{\,{\pm}\,}lr@{\,{\pm}\,}lr@{\,{\pm}\,}l}
\hline
\hline
Parameter                                                          &   \mctc{  Channel 1       }    &   \mctc{   Channel 2      }    &   \mctc{    Channel 4     }        \\
\hline
Array position (\math{\bar{x}}, pixel)                               &   \mctc{      14.16       }    &   \mctc{      23.82       }    &   \mctc{       24.6       }        \\
Array position (\math{\bar{y}}, pixel)                               &   \mctc{      15.69       }    &   \mctc{      24.11       }    &   \mctc{       21.9       }        \\
Position consistency\tablenotemark{a} (\math{\delta\sb{\rm x}}, pixel)   &   \mctc{      0.005       }    &   \mctc{      0.02        }    &   \mctc{      0.021       }        \\
Position consistency\tablenotemark{a} (\math{\delta\sb{\rm y}}, pixel)   &   \mctc{      0.012       }    &   \mctc{      0.014       }    &   \mctc{      0.025       }        \\
Aperture size (pixel)                                                &   \mctc{      2.75        }    &   \mctc{       2.5        }    &   \mctc{      3.5         }        \\
Sky Annulus inner radius (pixel)                                     &   \mctc{       8.0        }    &   \mctc{      12.0        }    &   \mctc{      12.0        }        \\
Sky Annulus outer radius (pixel)                                     &   \mctc{      20.0        }    &   \mctc{      30.0        }    &   \mctc{      30.0        }        \\
System flux \math{F\sb{\rm s}} (\micro Jy)                         &          102802 & 4            &         66083 & 7              &         24381 & 3              \\
Eclipse depth (\%)                                                 &         0.187 & 0.007          &         0.224 & 0.018          &         0.181 & 0.022              \\
Brightness temperature (K)                                         &         2225 &  39             &         2212  & 94             &        1590 & 116                  \\
Eclipse midpoint (orbits)                                          &        0.4825 & 0.0003         &        0.4842 & 0.0005         &        0.4842 & 0.0005             \\
Eclipse midpoint (BJD\sb{UTC} --2,450,000)                         &      5274.6609 & 0.0006        &      4908.9290 & 0.0011        &      4908.9290 & 0.0011            \\
Eclipse midpoint (BJD\sb{TDB} --2,450,000)                         &      5274.6617 & 0.0006        &      4908.9298 & 0.0011        &      4908.9298 & 0.0011            \\
Eclipse duration  (\math{t\sb{\rm 4-1}}, hrs)                      &          2.59 & 0.03           &         2.59 & 0.03            &         2.59 & 0.03                \\
Ingress/egress time (\math{t\sb{\rm{2-1}}}, hrs)                   &         0.290 & 0.007          &         0.290 & 0.007          &          0.290 & 0.007             \\
Ramp name                                                          &   \mctc{     linear       }    &   \mctc{       ...        }    &   \mctc{       ...        }        \\
Ramp, linear term (\math{r\sb{\rm 0}})                             &          0.0044 & 0.0010       &   \mctc{       ...        }    &   \mctc{       ...        }        \\
Intrapixel method                                                  &   \mctc{      BLISS       }    &   \mctc{     BLISS        }    &   \mctc{       ...        }        \\
BLISS bin size in $x$  (pixel)                                       &   \mctc{     0.004        }    &   \mctc{     0.028        }    &   \mctc{       ...        }        \\
BLISS bin size in $y$  (pixel)                                       &   \mctc{     0.01         }    &   \mctc{     0.023        }    &   \mctc{       ...        }        \\
Minimum number of points per bin                                   &   \mctc{        4         }    &   \mctc{        5         }    &   \mctc{       ...        }        \\
Total frames                                                       &   \mctc{      13693       }    &   \mctc{       2972       }    &   \mctc{       1432       }        \\
Rejected frames (\%)                                               &   \mctc{     0.49         }    &   \mctc{     0.34         }    &   \mctc{     3.89         }        \\
Free parameters                                                    &   \mctc{        6         }    &   \mctc{        3         }    &   \mctc{        2         }        \\
AIC value                                                          &   \mctc{  16695.8         }    &   \mctc{   16695.8        }    &   \mctc{     16695.8      }        \\
BIC value                                                          &   \mctc{  16780.7         }    &   \mctc{   16780.7        }    &   \mctc{     16780.7      }        \\
SDNR                                                               &   \mctc{  0.003311        }    &   \mctc{   0.004473       }    &   \mctc{    0.003980      }        \\
Uncertainty scaling factor                                         &   \mctc{  0.031968        }    &   \mctc{   0.294486       }    &   \mctc{     0.342520     }        \\
Photon-limited S/N (\%)                                            &   \mctc{    72.7          }    &   \mctc{    90.4          }    &   \mctc{       68.1       }        \\
\hline\end{tabular}
\vspace{-5pt}
\begin{minipage}[t]{0.65\linewidth}
\tablenotetext{1}{rms frame-to-frame position difference.}
\end{minipage}
\end{table*}
\if\submitms y
\clearpage
\fi


\begin{table*}[ht]
\vspace{-5pt}
\centering
\caption{\label{tab:eclfits-sin} Joint Best-fit Eclipse Light-curve Parameters (Channel 1--Sinusoidal Ramp)}
\begin{tabular}{lr@{\,{\pm}\,}lr@{\,{\pm}\,}lr@{\,{\pm}\,}l}
\hline
\hline
Parameter                                                          &   \mctc{   Channel 1  }        &   \mctc{   Channel 2   }      &   \mctc{    Channel 4     }     \\
\hline
Array position (\math{\bar{x}}, pixel)                               &   \mctc{      14.16       }    &   \mctc{      23.82       }    &   \mctc{       24.6       }      \\
Array position (\math{\bar{y}}, pixel)                               &   \mctc{      15.69       }    &   \mctc{      24.11       }    &   \mctc{       21.9       }      \\
Position consistency\tablenotemark{a} (\math{\delta\sb{x}}, pixel)   &   \mctc{      0.005       }    &   \mctc{       0.02       }    &   \mctc{      0.021       }       \\
Position consistency\tablenotemark{a} (\math{\delta\sb{y}}, pixel)   &   \mctc{      0.012       }    &   \mctc{      0.014       }    &   \mctc{      0.025       }       \\
Aperture size (pixel)                                                &   \mctc{      2.75        }    &   \mctc{      2.5         }    &   \mctc{      3.5         }      \\
Sky Annulus inner radius (pixel)                                     &   \mctc{      8.0         }    &   \mctc{     12.0         }    &   \mctc{     12.0         }       \\
Sky Annulus outer radius (pixel)                                     &   \mctc{     20.0         }    &   \mctc{     30.0         }    &   \mctc{     30.0         }      \\
System flux \math{F\sb{s}} (\micro Jy)                             &       102616 & 7               &         66083 & 7              &       24381 & 3             \\
Eclipse depth (\%)                                                 &         0.193 & 0.007          &         0.224 & 0.017          &         0.181 & 0.021              \\
Brightness temperature (K)                                         &         2258 &  38             &         2212  & 89             &        1590 & 111                  \\
Eclipse midpoint (orbits)                                          &         0.4825 & 0.0003        &          0.4843 & 0.0005       &        0.4843 & 0.0005           \\
Eclipse midpoint (BJD\sb{UTC} --2,450,000)                         &      5274.6609 & 0.0006        &       4908.9291 & 0.0011        &     4908.9291 & 0.0011           \\
Eclipse midpoint (BJD\sb{TDB} --2,450,000)                         &      5274.6617 & 0.0006        &       4908.9298 & 0.0011        &     4908.9298 & 0.0011           \\
Eclipse duration (\math{t\sb{\rm{4-1}}}, hrs)                      &           2.59 & 0.03          &            2.59 & 0.03         &          2.59 & 0.03            \\
Ingress/egress time (\math{t\sb{\rm{2-1}}}, hrs)                   &          0.290 & 0.007         &           0.290 & 0.007        &         0.290 & 0.007            \\
Ramp name                                                          &   \mctc{     sinusoidal   }    &   \mctc{      ...          }   &   \mctc{       ...         }     \\
Ramp, cosine phase offset (\math{t\sb{2}})                         &           0.5356 & 0.0016      &   \mctc{      ...           }   &   \mctc{      ...         }      \\
Intrapixel method                                                  &   \mctc{      BLISS       }    &   \mctc{     BLISS         }   &   \mctc{      ...         }      \\
BLISS bin size in $x$  (pixel)                                       &   \mctc{     0.004        }    &   \mctc{     0.028        }    &   \mctc{       ...         }        \\
BLISS bin size in $y$  (pixel)                                       &   \mctc{     0.01         }    &   \mctc{     0.023        }    &   \mctc{      ...         }        \\
Minimum number of points per bin                                   &   \mctc{        4         }    &   \mctc{        5          }   &   \mctc{        ...        }      \\
Total frames                                                       &   \mctc{      13693       }    &   \mctc{       2972        }   &   \mctc{       1432       }      \\
Rejected frames (\%)                                               &   \mctc{    0.49          }    &   \mctc{      0.34         }   &   \mctc{    3.89          }      \\
Free parameters                                                    &   \mctc{        6         }    &   \mctc{        3          }   &   \mctc{        2         }      \\
AIC value                                                          &   \mctc{  16695.9         }    &   \mctc{  16695.9          }   &   \mctc{  16695.9         }      \\
BIC value                                                          &   \mctc{  16780.8         }    &   \mctc{  16780.8          }   &   \mctc{  16780.8         }      \\
SDNR                                                               &   \mctc{ 0.003316         }    &   \mctc{ 0.004473          }   &   \mctc{ 0.003980         }      \\
Uncertainty scaling factor                                         &   \mctc{  0.031968        }    &   \mctc{  0.294485         }   &   \mctc{  0.342520        }      \\
Photon-limited S/N (\%)                                            &   \mctc{       72.6       }    &   \mctc{       90.4        }   &   \mctc{       68.1       }      \\
\hline\end{tabular}
\vspace{-5pt}
\begin{minipage}[t]{0.65\linewidth}
\tablenotetext{1}{rms frame-to-frame position difference.}
\end{minipage}
\end{table*}
\if\submitms y
\clearpage
\fi

\section{ORBIT}
\label{sec:orbit}

We fit the midpoint times from the {\em Spitzer}\/ lightcurves simultaneously with
the available radial velocity curves and transit photometry in order to
provide updated estimates of system orbital parameters. The timing of secondary eclipse is a strong constraint on the shape and orientation of the
orbit. The two eclipses for the linear and sinusoidal joint fit (Tables\ \ref{tab:eclfits-lin} and \ref{tab:eclfits-sin}) have an insignificant difference in phases (less than 0.5\math{\sigma}), and the linear joint fit has slightly lower BIC value. Hence, we picked the linear joint fit phases for the use in the orbital analysis. The two eclipses occur at phases 0.4825 {\pm} 0.0003 and  0.4841 {\pm} { 0.0005} (using the \citealp{Joshi2009-WASP14b} ephemeris), with a weighted mean after a 37 s eclipse-transit light-time correction of 0.48273 {\pm} 0.00025, indicating that \math{e \cos \omega} = - 0.0271 {\pm} 0.0004. The phases differ from each other by approximately 3\math{\sigma}, but depend strongly on the accuracy of the ephemeris used to compute them. 

We fit a Keplerian orbit model to our secondary eclipse times along with
radial velocity data from \cite{Husnoo2011-WASP14b} and
\cite{Joshi2009-WASP14b}, and transit timing data from both amateur
observers and WASP-14b's discovery paper
\citep{Joshi2009-WASP14b}.  The entire data set comprised 38 RV points, six
of which were removed due to the Rossiter-McLaughlin effect, 30 transits,
and two eclipses (see Table\ \ref{tab:ttv}). All times were adjusted to BJD\sb{TDB}
\citep{EastmanEtal2010apjLeapSec}. The errors were estimated using
our MCMC routine.   This fit gave \math{e = 0.087 \pm
 0.002} and \math{\omega = 107{\degrees}.1 {\pm} 0{\degrees}.5}. We did not adjust for any anomalous eccentricity signal from the stellar tidal bulge as described by \citet{ArrasEtal2012MNRAS-RVstar} because the predicted amplitude of this effect is smaller than the uncertainty on the eccentricity, and much smaller than the eccentricity itself. With our new data, we
refine the ephemeris to \math{T\sb{\rm BJD\sb{\rm TDB}} = 2454827.06666(24) + 
2.2437661(11)\,N}, where \math{T} is the time of transit and \math{N} is the
number of orbits elapsed since the transit time (see Table\ \ref{tab:orbit}).  We find that the new
ephemeris reduces the difference between the two eclipse phases to less than
1.6\math{\sigma}.  Performing an ephemeris fit to the transit and eclipse data
separately shows that the transit and eclipse periods differ by (1.1 \pm
{ 0.8}) \math{\times} 10\sp{-5} days, a 1.5\math{\sigma} result that limits apsidal motion,
\math{\dot{\omega}}, to less than 0{\degree}.0024 day\sp{-1} at the 3\math{\sigma} level
\citep{GimenezBastero1995}. 

The results confirm an eccentric orbit for WASP-14b and improve knowledge of
other orbital parameters.

\section{ATMOSPHERE}
\label{sec:atm}

\begin{table}[ht]
\vspace{-5pt}
\centering
\caption{\label{tab:ttv} Transit Timing Data}
\begin{tabular}{lcl}
\hline
\hline
Mid-transit Time (BJD\sb{TDB})  &  Uncertainty          &   Source\tablenotemark{a}                                     \\
\hline
2455695.4082	                & 0.0012	        & V.\ Slesarenkno, E.\ Sokov \tablenotemark{b}                    \\
2455668.4790	                & 0.0011		& Franti\^{s}ek Lomoz						\\
2455652.7744	                & 0.0014		& Stan Shadick, C.\ Shiels\tablenotemark{c}		        \\
2455650.5307	                & 0.0018		& Lubos Br\'{a}t						\\
2455650.52789	                & 0.00076	        & Martin Vra\v{s}t'\'{a}k					\\
2455650.52566	                & 0.00067		& Jaroslav Trnka\tablenotemark{d}				\\
2455632.5807	                & 0.0011		& E.\ Sokov, K.\ N.\ Naumov \tablenotemark{b}                      \\
2455318.45101	                & 0.00085		& Anthony Ayiomamitis						\\
2455302.7464	                & 0.0010		& Stan Shadick\tablenotemark{c}					\\
2455264.6021	                & 0.0012		& Hana Ku\v{c}\'{a}kov\'{a}\tablenotemark{e}			\\
2455264.6017	                & 0.0013		& Radek Koci\'{a}n\tablenotemark{f}				\\
2455219.7290	                & 0.0012		& Lubos Br\'{a}t						\\
2454979.643		        & 0.003			& Wiggins, AXA						        \\
2454968.426		        & 0.001			& Srdoc, AXA						        \\
2454950.4831	                & 0.0021		& Jesionkiewicz, AXA						\\
2454950.4746                	& 0.0014		& Lubos Br\'{a}t						\\
2454950.4745	                & 0.0018		& Hana Ku\v{c}\'{a}kov\'{a}\tablenotemark{e}			\\
2454950.4731	                & 0.0021		& Pavel Marek						        \\
2454950.4728	                & 0.0014		& Wardak, AXA						        \\
2454943.7427	                & 0.0006		& Dvorak, AXA						        \\
2454941.49799	                & 0.00081		& Jaroslav Trnka\tablenotemark{d}				\\
2454941.4916	                & 0.0019		& Franti\v{s}ek Lomoz						\\
2454934.765		        & 0.001			& Brucy Gary, AXA						\\
2454932.5246	                & 0.0014		& Radek D\v{r}ev\v{e}n\'{y}					\\
2454932.5232	                & 0.0011		& Lubos Br\'{a}t						\\
2454932.5222	                & 0.0013		& Jaroslav Trnka\tablenotemark{d}				\\
2454932.5219	                & 0.0015		& T.\ Hynek, K.\ Onderkov\'{a}					\\
2454914.5753	                & 0.0008		& Naves, AXA						        \\
2454887.6457	                & 0.0014		& Georgio, AXA						        \\
\hline
\end{tabular}
\begin{minipage}[t]{0.95\linewidth}
\tablenotetext{1}{The Amateur Exoplanet Archive (AXA), http://brucegary.net/AXA/x.htm) and Transiting ExoplanetS and Candidates group (TRESCA), http://var2.astro.cz/EN/tresca/index.php) supply their data to the Exoplanet Transit Database (ETD), http://var2.astro.cz/ETD/), which performs the uniform transit analysis described by \citet{Poddany2010}. The ETD Web site provided the AXA and TRESCA numbers in this table, which were converted to BJD\sb{TDB}.}
\tablenotetext{2}{Sokov E., Naumov K., Slesarenko V.\ et al., Pulkovo Observatory of RAS, Saint-Petersburg, Russia.}
\tablenotetext{3}{Physics and Engineering Physics Department, University of Saskatchewan, Saskatoon, Saskatchewan, Canada, S7N 5E2.}
\tablenotetext{4}{Municipal Observatory in Slany Czech Republic.}
\tablenotetext{5}{Project Eridanus, Observatory and Planetarium of Johann Palisa in Ostrava.}
\tablenotetext{6}{Koci\'{a}n R., Johann Palisa, Observatory and Planetarium, Technical University Ostrava, 17.\ Listopadu 15, CZ-708 33 Ostrava, Czech Republic.}
\end{minipage}
\vspace{-13pt}
\end{table}
\if\submitms y
\clearpage
\fi

\begin{table}[h]
\centering
\caption{\label{tab:orbit} Eccentric Orbital Model}
\begin{tabular}{lr@{\,{\pm}\,}lr@{\,{\pm}\,}}
\hline
\hline 
Parameter                                                          &  \mctc{Value}                       \\
\hline
\math{e \sin \omega}\tablenotemark{a}     	                   &  0.0831               & 0.0021      \\
\math{e \cos \omega}\tablenotemark{a}     	                   &  \math{-0.02557}      & 0.00038     \\
\math{e}                          			           &  0.087                & 0.002       \\
\math{\omega} (\degree)           			           &  \math{-107.1}        & 0.5         \\
\math{P} (days)\tablenotemark{a}      		                   &  2.2437661            & 0.0000011   \\
\math{T\sb{\rm 0}} \tablenotemark{a}\tablenotemark{b}              &  2454827.06666        & 0.00024     \\
\math{K} (m\,s\sp{-1})\tablenotemark{a}\tablenotemark{c}           &  990                  & 3           \\
\math{\gamma} (m\,s\sp{-1})\tablenotemark{a}\tablenotemark{d}      &  \math{-4987.9}      & 1.6         \\
\math{\chi^2}				                           &  \mctc{162}                         \\
\hline                                           
\end{tabular}
\begin{minipage}[t]{0.55\linewidth}
\tablenotemark{a}{Free parameter in MCMC fit.}\\
\tablenotemark{b}{BJD\sb{TDB}.}\\
\tablenotemark{c}{Radial velocity semi-amplitude.}\\
\tablenotemark{d}{Radial velocity offset.}
\end{minipage}
\end{table}
\if\submitms y
\clearpage
\fi

We explore the model parameter space in search of the best-fitting models for a given data set. The model parameterization is described by \citet{MadhusudhanSeager2009,MadhusudhanSeager2010, Madhusudhan2012}. The sources of opacity in the model include molecular absorption due to H\sb{2}O, CO, CH\sb{4}, CO\sb{2}, TiO, and VO, and collision-induced absorption (CIA) due to H\sb{2}-H\sb{2}. Our molecular line lists are obtained from \citet{Freedman08}, R. S. Freedman (2009, private communication), \citet{Rothman2005-HITRAN, KarkoschkaTomasko2010}, and E. Karkoschka (2011, private communication). Our CIA opacities are obtained from \citet{Borysow1997} and \citet{Borysow2002}. We explore the model parameter space using a MCMC scheme, as described by \citet{MadhusudhanSeager2010}. However, since the number of model parameters (\math{n} = 10) exceed the number of data points (\math{N\sb{\rm{data}}} = 3), our goal is not to find a unique fit to the data but, primarily, to identify regions of model phase space that the data exclude. In order to compute the model planet-star flux ratios to match with the data, we divide the planetary spectrum by a Kurucz model of the stellar spectrum derived from \citet{Castelli2004}. Our models allow constraints on the temperature structure, molecular mixing ratios, and a joint constraint on the albedo and day-night redistribution. 

\begin{figure}[ht]
    \centering
    \includegraphics[width=0.47\textwidth, clip]{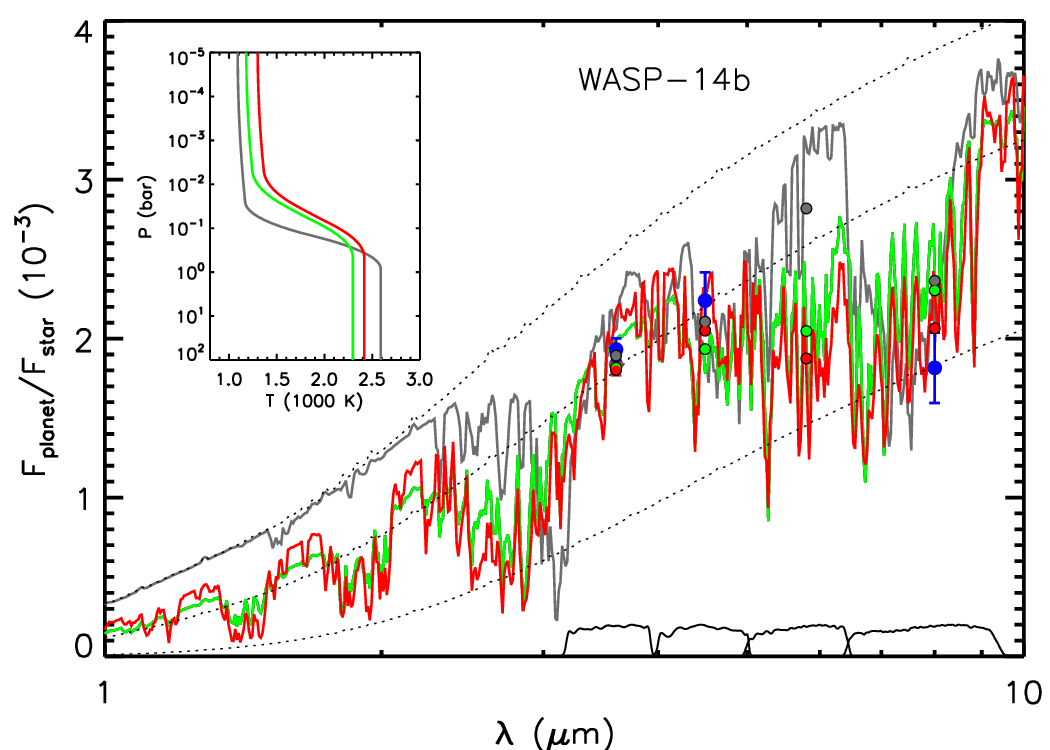}
\caption {\footnotesize {\label{fig:spectra}
Observations and model spectra for dayside emission from WASP-14b. The blue filled circles with error bars show our observations in {\em Spitzer} channel 1 (3.6 {\micron}), 2 (4.5 {\micron}), and 4 (8.0 {\micron}). For the 3.6 {\micron} channel, two values are shown, in blue and brown, corresponding to different ramp models used in deriving the eclipse depths (see Section\ \ref{sec:ch1}). The green, red, and gray curves show model spectra with different chemical compositions and without thermal inversions that explain the data; the corresponding pressure--temperature (\math{P-T}) profiles are shown in the inset.  The green model has molecular abundances in thermochemical equilibrium assuming solar elemental abundances. The red model has 10 times lower CO and 6 times higher H\sb{2}O compared to solar abundance chemistry, i.e., more oxygen-rich than solar abundances. The gray model has a carbon-rich chemistry (C/O = 1). The green, red, and gray circles show the model spectra integrated in the {\em Spitzer} IRAC bandpasses. The oxygen-rich (red) model provides a marginally better fit to the data than the solar and carbon-rich models. The black dotted lines show three blackbody planet spectra at 1600 K, 2200 K, and 2600 K}.
}
\end{figure}

We find that strong constraints can be placed on the presence of a thermal inversion in WASP-14b even with our current small set of observations. At an irradiation of  3 \math{\times} 10\sp{9} erg\,s\sp{-1}\,cm\sp{-2}, WASP-14b falls in the class of extremely irradiated planets that are predicted to host thermal inversions according to the TiO/VO hypothesis of \citet{Fortney2008}. However, the present observations do not show any distinct evidence of a thermal inversion in the dayside atmosphere of WASP-14b. We explored the model parameter space by running \sim 10\sp{6} models with and without thermal inversions, using an MCMC scheme as discussed above. We found that the data could not be explained by a thermal inversion model for any chemical composition. On the other hand, the data are easily fit by models with no thermal inversions. While the brightness temperatures in the 3.6 and 4.5 {\micron} channels are consistent with a blackbody spectrum of the planet at \math{T}\sim 2200 K, the 8 {\micron} flux deviates substantially from the assumption of a blackbody with a brightness temperature of 1668 {\pm} 125 K. In the presence of a thermal inversion, the flux in the 8 {\micron} channel is expected to be much higher than the fluxes in the 3.6 and 4.5 {\micron} channels due to emission features of water vapor and, if present, methane. The low flux observed at 8 {\micron}, therefore, implies strong water vapor and/or methane in absorption, implying the lack of a significant temperature inversion (see \citealp{MadhusudhanSeager2010} for a discussion on inferring thermal inversions). We also note that, as mentioned in Section\ \ref{sec:ch1}, the observations yield different planet--star flux contrasts in the 3.6 {\micron} channel for different choices of ramp models. However, as shown in Figure \ref{fig:spectra}, the two extreme values are still consistent at the 1\math{\sigma} level, and as such, lead to similar model conclusions.


We modeled the dayside atmosphere of WASP-14b using the exoplanetary atmospheric modeling method developed by \citet{MadhusudhanSeager2009,MadhusudhanSeager2010}. We use a one-dimensional line-by-line radiative transfer code to model the planetary atmosphere under the assumption of local thermodynamic equilibrium, hydrostatic equilibrium, and global energy balance at the top of the atmosphere. The latter condition assumes that the integrated emergent planetary flux balances the integrated incident stellar flux, accounting for the Bond albedo (\math{A\sb{B}}) and possible redistribution of energy onto the night side. Our model uses parameterized prescriptions to retrieve the temperature structure and chemical composition from the observations, as opposed to assuming radiative and chemical equilibrium  with fixed elemental abundances \citep{BurrowsEtal2008apjSpectra, Fortney2008}.


\begin{figure*}[ht]
    \centering
    \includegraphics[width=0.80\textwidth, clip]{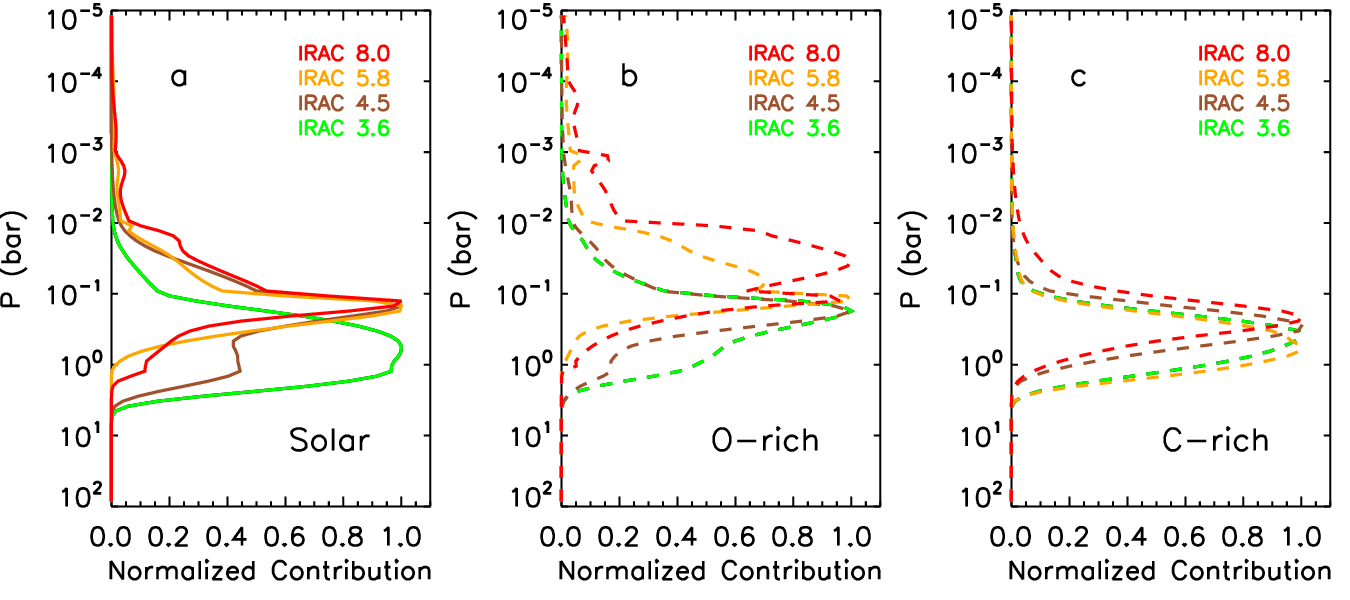}
\caption {\footnotesize {\label{fig:contr-func}
Left: contribution functions in the four {\em Spitzer} channels corresponding to the green model shown in Figure \ref{fig:spectra}. The legend shows the channel center wavelength in {\microns} and the curves are color-coded by the channel. All the contribution functions are normalized to unity.
Middle: contribution functions corresponding to the red model shown in Figure \ref{fig:spectra}.
Right: contribution functions corresponding to the gray model shown in Figure \ref{fig:spectra}}.
}
\end{figure*}

We find that the data can be explained by models with a wide range of chemical compositions. Figure \ref{fig:spectra} shows three model spectra with different chemistries, along with the observations: (1) a solar-abundance model (in green in Figure \ref{fig:spectra}) with chemical composition in thermochemical equilibrium assuming solar abundances (TE\sb{\rm solar}), (2) an oxygen-rich model (in red) with 10{\by} lower CO and 8{\by} higher H\sb{2}O, and (3) a carbon-rich model (in gray, e.g., \citealp{MadhusudhanEtal2011natWASP12batm}, \citealp{Madhusudhan2012}). The oxygen-rich model fits the data marginally better than the solar abundance model. A slightly lower CO is favored because of the slightly higher 4.5 {\micron} flux compared to the 3.6 {\micron} flux, which means lower absorption due to CO. Higher absorption due to H\sb{2}O is favored by the low 8 {\micron} point. In principle, a lower CO and a higher H\sb{2}O, compared to TE\sb{\rm solar} values, are both possible by having a C/O ratio less than the solar value of 0.54. However, more data would be required to confirm the low CO requirement, because a blackbody of \sim 2200 K fits the 3.6 and 4.5 {\micron} points just as well.

Models with high C/O ratios (C/O \math{\geq} 1, i.e., carbon-rich), can lead to strong CH\sb{4}, C\sb{2}H\sb{2}, and HCN absorption in the 3.6 {\micron} and 8 {\micron} channels \citep[e.g.,][]{MadhusudhanEtal2011natWASP12batm, Madhusudhan2011b, Madhusudhan2012}, instead of H\sb{2}O absorption in the low-C/O models. As shown in Figure \ref{fig:spectra}, the C-rich model fits the data as well as the solar-abundance model, but less precisely than the model with low C/O (i.e., enhanced H\sb{2}O and low CO). Although the data marginally favor an oxygen-rich composition in the dayside atmosphere of WASP-14b, new observations are required to provide more stringent constraints on the C/O ratio. Future observations in the near-infrared, from ground and space, can place further constraints on the temperature structure and composition, especially the C/O ratio, of the dayside atmosphere of WASP-14b. In particular, as shown in Figure \ref{fig:spectra}, near-infrared observations in the 1--2.5 {\micron} range probe spectral features of several oxygen- and carbon-bearing molecules such as H\sb{2}O, CO, and CH\sb{4}, mixing ratios of which can provide stringent constraints on the C/O ratio \citep{Madhusudhan2011b}. For example, the oxygen-rich models predict deep absorption features in the H\sb{2}O bands, contrary to the carbon-rich model, which contains no significant water absorption.  {\em Hubble Space Telescope} WFC3 observations in the 1.1--1.7 {\micron} range can test for water absorption. Furthermore, the models with different C/O ratios also predict different continuum fluxes, which can be observed from ground in the \math{J}, \math{H}, and \math{K} bands \citep[see][]{Madhusudhan2012}.

The models explaining the observations require relatively low day-night energy redistribution in WASP-14b. As shown by the contribution functions in Figure \ref{fig:contr-func}, the 3.6 {\micron} IRAC channel probes the atmosphere between 0.1 and 1 bar.  Consequently, the high brightness temperature in the 3.6 {\micron} channel indicates a hot planetary photosphere. Over the entire model population explored by our retrieval method, we find that the data allow for up to \sim 30\% of the energy incident on the dayside to be redistributed to the night side (i.e., for zero Bond albedo). For the particular best-fitting model (in red) shown in Figure \ref{fig:spectra}, this fraction is \sim 25\%. For non-zero albedos the fraction is even lower; since the quantity we constrain is \math{\eta} = (1-\math{A}\sb{\rm B})(1-\math{f}\sb{\rm r}), where \math{A}\sb{\rm B} is the Bond albedo and \math{f}\sb{\rm r} is the fraction of the dayside incident energy redistributed to the nightside \citet{MadhusudhanSeager2009}. However, the present constraints on the day-night redistribution are only suggestive and new observations are essential to further constrain the energy redistribution in WASP-14b. For example, observations in atmospheric windows at lower wavelengths, e.g., between 1 and 2 {\micron}, where the black-body of the planetary photosphere would peak, would be critical to further constrain the lower atmospheric thermal structure, and hence the energy budget of the planet's dayside atmosphere. More importantly, phase-curve observations are required to constrain the day-night energy redistribution directly \citep[e.g.,][]{Knutson2007, KnutsonEtal2009ApJ-Tres4Inversion}.

\section{DISCUSSION}
\label{sec:discus}

The absence of a thermal inversion in the dayside atmosphere of WASP-14b constrains inversion-causing phenomena in irradiated atmospheres. The canonical argument for such inversions is via absorption in the optical by gaseous TiO and VO \citep{Hubeny2003, Fortney2008}. On the other hand, \citet{Spiegel2009} showed that the high mean molecular masses of TiO and VO would lead to significant gravitational settling of these molecules, thereby depleting them from the upper atmospheres, unless strong vertical mixing keeps them aloft. Additionally, the abundances of inversion-causing molecules might also be influenced by stellar activity and photochemistry \citep{KnutsonHowardIsaacson2010ApJ-CorrStarPlanet}. Consequently, the real cause of thermal inversions in irradiated atmospheres is currently unknown. Nevertheless, models used to infer thermal inversions in the literature have either used parameterized visible opacity sources \citep{BurrowsEtal2008apjSpectra} or parametric temperature profiles (\citealp{MadhusudhanSeager2009}, also used in the present work). To first order, the lack of a thermal inversion in WASP-14b might indicate that the vertical mixing in the dayside atmosphere of WASP-14b is weaker compared to the downward diffusion of TiO and VO.

\begin{table*}[ht]
\caption{\label{table:SystemParams} 
System Parameters of WASP-14}
\atabon\strut\hfill\begin{tabular}{llc}
    \hline
    \hline
    Parameter                                                              & Value                   & Reference                 \\
    \hline

    \multicolumn{3}{c}{Orbital parameters}                                                                                       \\ \hline

    Orbital period, \math{P} (days)                                        &    2.2437661 {\pm} 0.0000011    & a                \\
    Semimajor axis, \math{a} (AU)                                          &    0.036 {\pm} 0.001            & b                \\
    Transit time (BJD\sb{TDB})                                             &    2454827.06666 {\pm} 0.00024  & a                \\
    Orbital eccentricity, \math{e}                                         &    0.087 {\pm} 0.002            & a                \\
    Argument of pericenter, \math{\omega} (deg)                            &    \math{-107.1} {\pm} 0.5      & a                \\
    Velocity semiamplitude, \math{K} (m\,s\sp{-1})                         &    990.0 {\pm} 3                & a                \\ 
    Centre-of-mass velocity \math{\gamma} (m\,s\sp{-1})                    &    \math{-4987.9} {\pm} 1.6     & a                \\ \hline
 
    \multicolumn{3}{c}{Stellar parameters}                                                                                       \\ \hline

    Spectral type                                                          &    F5V                           & b                \\
    Mass, \math{M\sb{\rm *}} (\math{M\sb{\odot}})                          &    1.211 $^{+0.127}_{-0.122}$    & b                \\
    Radius, \math{R\sb{\rm *}} (\math{R\sb{\odot}})                        &    1.306 $^{+0.066}_{-0.073}$    & b                \\
    Mean density, \math{\rho\sb{\rm *}} (\math{\rho\sb{\odot}})            &    0.542 $^{+0.079}_{-0.060}$    & b                \\
    Effective temperature, \math{T\sb{\rm eff}} (K)                        &    6475 {\pm} 100                & b                \\
    Surface gravity, log \math{g\sb{\rm *}} (cgs)                          &    4.287 $^{+0.043}_{-0.038}$    & b                \\
    Projected rotation rate, \math{v\sb{\rm *} \sin(i)} (kms\sp{-1})       &    4.9 {\pm} 1.0                 & b                \\
    Metallicity [M/H] (dex)                                                &    0.0 {\pm} 0.2                 & b                \\
    Age (Gyr)                                                              &    \sim 0.5--1.0                 & b                \\
    Distance (pc)                                                          &    160 {\pm} 20                  & b                \\ 
    Lithium abundance, log \math{N}(Li)                                    &    2.84 {\pm} 0.05               & b                \\ \hline

    \multicolumn{3}{c}{Planetary parameters }                                                                                    \\ \hline

    Transit depth, (\math{R\sb{\rm p}}/\math{R\sb{\rm star}})\sp2          &    0.0102 $^{+0.0002}_{-0.0003}$ & b               \\
    Mass, \math{M\sb{\rm p}} (\math{M\sb{\rm J}})                          &    7.341 $^{+0.508}_{-0.496}$    & b               \\
    Radius, \math{R\sb{\rm p}} (\math{R\sb{\rm J}})                        &    1.281 $^{+0.075}_{-0.082}$    & b               \\
    Surface gravity, log \math{g\sb{\rm p}} (cgs)                          &    4.010 $^{+0.049}_{-0.042}$    & b               \\
    Mean density, \math{{\rho}\sb{\rm p}} (g\,cm\sp{-3})                   &    4.6                           & b               \\
    Equilibrium temperature (\math{A}=0), \math{T\sb{\rm eq}} (K)          &    1866.12 $^{+36.74}_{-42.09}$  & b               \\
    \hline
    \multicolumn{2}{l}{\sp{a}  Our analyses (see Section\ \ref{sec:orbit})}\\
    \multicolumn{2}{l}{\sp{b} \citet{Joshi2009-WASP14b}}
\end{tabular}\hfill\strut\ataboff
\end{table*}

{\em Spitzer} has observed a number of strongly irradiated hot Jupiters with brightness temperatures in the 1000--2000 K range. The inferences of thermal inversions from emission photometry result from flux excesses in molecular bands where strong absorption is expected \citep{MadhusudhanSeager2010}. In principle, detection of a thermal inversion is possible with just two {\em Warm Spitzer}\/ channels with sufficient S/N if there is a large flux difference between channels 1 and 2 \citep{Knutson08, KnutsonEtal2009ApJ-Tres4Inversion, MadhusudhanSeager2010, MachalekEtal2009ApJ-XO2b, O'DonovanEtal2010ApJ-SpitzerTres2, CristiansenEtal2010ApJ-HATP7b}. Based on the TiO/VO hypothesis described above, \citet{Fortney2008} suggested that depending on the level of irradiation from their parent star, irradiated planets can fall into two categories: the very highly irradiated atmospheres that host thermal inversions and the less-irradiated ones that do not.  However, recent observations have revealed several counterexamples to this hypothesis. \citet{Machalek2008-XO-1b} present evidence for a temperature inversion in XO-1b, despite low irradiation of the planet (\math{T\sb{\rm eq}} = 1209 K), while \citet{FressinEtal2010ApJ-Tres3} show no thermal inversion, although TrES-3b is a highly irradiated planet (\math{T\sb{\rm eq}} = 1643 K). Similarly, WASP-12b, one of the most irradiated hot Jupiters known, has also been reported to lack a significant thermal inversion \citep{MadhusudhanEtal2011natWASP12batm}. In this paper, we present WASP-14b as another counterexample. It is possible that additional parameters (e.g., metallicity, surface gravity, C/O ratio) influence the presence or the absence of a temperature inversion. However, more observations are needed to explain WASP-14b's missing inversion.

\section{CONCLUSIONS}
\label{sec:concl}

During two secondary eclipse events, {\em Spitzer}\/ observed WASP-14b in three IRAC channels: 3.6, 4.5, and 8.0 {\micron}. All eclipses have a high S/N (3.6 {\micron} channel \sim 25, 4.5 {\micron} channel \sim 12, 8.0 {\micron} channel \sim 8), which allowed us to constrain the planetary spectrum and orbital parameters. 

Our observations probe the atmosphere at pressures between 0.01 and 1 bar and indicate the absence of a significant thermal inversion in the dayside atmosphere of WASP-14b. Given WASP-14b's highly irradiated atmosphere, this contradicts predictions that the most-irradiated hot Jupiters should have thermal inversions due to gaseous TiO/VO \citep{Fortney2008}. Additionally, our observations place nominal constraints on the chemical composition and day-night energy redistribution in the atmosphere of WASP-14b. We find that the data can be explained by non-inversion models with nearly solar abundances in chemical equilibrium. A factor of 10 less CO and a factor of 6 higher H\sb{2}O, compared to those obtained with solar abundances, explain the data to within the 1\math{\sigma} uncertainties, on average. Such CO depletion and H\sb{2}O enhancement are, in principle, possible in chemical equilibrium with C/O ratios lower than solar. More data are required to constrain the atmospheric composition of WASP-14b better. 

Because the planet is much brighter than its predicted equilibrium temperature for uniform redistribution (\math{T\sb{\rm eq}} = 1866 K), the best-fitting models limit day-night energy redistribution in WASP-14b to \math{\leq} 30\% for zero Bond albedo.  Thermal phase-curve observations can probe the nightside emission directly and better constrain this quantity.

WASP-14b is one of the most massive transiting planets known, along with CoRoT-3b \citep{Triaud2009-CoRoT3b, Deleuil2008-CoRoT3b}, HAT-P-2b \citep{Bakos2007-HATP-2b, Winn2007-HATP2b, Loeillet2008-HATP-2b}, XO-3b \citep{Hebrard2008-XO3b, Johns-Krull2008-XO-3b, winn:2008}, and WASP-18b \citep{NymeyerEtal2011apjWASP18b}. With the exception of WASP-18b, all of these objects have very eccentric orbits. Classically, closer planets should have more circular orbits due to greater tidal orbital decay. At distances \math{a} < 0.1 AU, circularization should occur in typically a few Myr, compared to common system ages of a few Gyr. However, \citet{Pont2011-mass-period} argue that the time to circularize scales with the planet-star mass ratio, and is also a steep function of the orbital separation scaled to the planet radius (see their Figure 3). For planets with \math{M > M}\sb{\rm J}, the mass-period relation (see their Figure 2) suggests that heavier planets get circularized very close to their parent star, or may not ever reach circularization in their lifetime. A possible explanation is that the planet raises tides on its host star strong enough that the angular momentum of the planet is transferred to the stellar spin, and the planet gets swallowed by the star. This does not oppose the classical tide theory \citep[e.g.,][]{Goldreich1966}, but rather suggests that stopping mechanisms and tidal circularization are related. WASP-14b also has unusually high density for a hot Jupiter, similar to that of some rocky planets (4.6 g\,cm\sp{-3}). The planet's strong signal makes it ideal for further observation to constrain its composition and thus possible formation mechanisms for it and similar objects.

\acknowledgments
We thank Heather Knutson for providing the 3.6 {\micron} {\em Spitzer}\/ data prior to their public release, and Andrew Collier Cameron for useful discussions. We also thank contributors to SciPy, Matplotlib, and the Python Programming Language; other contributors to the free and open-source community; the NASA Astrophysics Data System; and the JPL Solar System Dynamics group for free software and services.  This work is based on observations made with the {\em Spitzer Space Telescope}, which is operated by the Jet Propulsion Laboratory, California Institute of Technology under a contract with NASA. NASA provided support for this work through an award issued by JPL/Caltech and Astrophysics Data Analysis Program grant NNX13AF38G. NM acknowledges support from the Yale Center for Astronomy and Astrophysics through the YCAA postdoctoral Fellowship. JB was partially supported by NASA Earth and Space Sciences Fellowship NNX12AL83H.

{\em Facility:Spitzer}

\begin{appendix}

\section{System Parameters}
\label{sec:app}

Table \ref{table:SystemParams} lists WASP-14 system parameters derived from our analysis and the literature. The eclipse parameters are listed in Tables \ref{tab:eclfits-lin} and \ref{tab:eclfits-sin}.

\end{appendix}

\bibliography{wasp14b}

\begin{thebibliography}{78}
\expandafter\ifx\csname natexlab\endcsname\relax\def\natexlab#1{#1}\fi

\bibitem[{{Anderson} {et~al.}(2010){Anderson}, {Hellier}, {Gillon}, {Triaud},
  {Smalley}, {Hebb}, {Collier Cameron}, {Maxted}, {Queloz}, {West}, {Bentley},
  {Enoch}, {Horne}, {Lister}, {Mayor}, {Parley}, {Pepe}, {Pollacco},
  {S{\'e}gransan}, {Udry}, \& {Wilson}}]{AndersonEtal010ApJ-WASP17b}
{Anderson}, D.~R., {Hellier}, C., {Gillon}, M., {et~al.} 2010, \apj, 709, 159

\bibitem[{{Anderson} {et~al.}(2011){Anderson}, {Smith}, {Lanotte}, {Barman},
  {Collier Cameron}, {Campo}, {Gillon}, {Harrington}, {Hellier}, {Maxted},
  {Queloz}, {Triaud}, \& {Wheatley}}]{Anderson2011-Ch24-WASP17b}
{Anderson}, D.~R., {Smith}, A.~M.~S., {Lanotte}, A.~A., {et~al.} 2011, \mnras,
  416, 2108

\bibitem[{{Arras} {et~al.}(2012){Arras}, {Burkart}, {Quataert}, \&
  {Weinberg}}]{ArrasEtal2012MNRAS-RVstar}
{Arras}, P., {Burkart}, J., {Quataert}, E., \& {Weinberg}, N.~N. 2012, \mnras,
  422, 1761

\bibitem[{{Bakos} {et~al.}(2007){Bakos}, {Kov{\'a}cs}, {Torres}, {Fischer},
  {Latham}, {Noyes}, {Sasselov}, {Mazeh}, {Shporer}, {Butler}, {Stefanik},
  {Fern{\'a}ndez}, {Sozzetti}, {P{\'a}l}, {Johnson}, {Marcy}, {Winn}, {Sip{\H
  o}cz}, {L{\'a}z{\'a}r}, {Papp}, \& {S{\'a}ri}}]{Bakos2007-HATP-2b}
{Bakos}, G.~{\'A}., {Kov{\'a}cs}, G., {Torres}, G., {et~al.} 2007, \apj, 670,
  826

\bibitem[{{Balachandran}(1995)}]{Balachandran1995}
{Balachandran}, S. 1995, \apj, 446, 203

\bibitem[{{Ballard} {et~al.}(2010){Ballard}, {Charbonneau}, {Deming},
  {Knutson}, {Christiansen}, {Holman}, {Fabrycky}, {Seager}, \&
  {A'Hearn}}]{BallardEtalr2010PASP-NewIntrapixel}
{Ballard}, S., {Charbonneau}, D., {Deming}, D., {et~al.} 2010, \pasp, 122, 1341

\bibitem[{{Boesgaard} \& {Tripicco}(1986)}]{BoesgaardTripicco1986}
{Boesgaard}, A.~M. \& {Tripicco}, M.~J. 1986, \apj, 303, 724

\bibitem[{{Borysow}(2002)}]{Borysow2002}
{Borysow}, A. 2002, \aap, 390, 779

\bibitem[{{Borysow} {et~al.}(1997){Borysow}, {Jorgensen}, \&
  {Zheng}}]{Borysow1997}
{Borysow}, A., {Jorgensen}, U.~G., \& {Zheng}, C. 1997, \aap, 324, 185

\bibitem[{{Burrows} {et~al.}(2008){Burrows}, {Budaj}, \&
  {Hubeny}}]{BurrowsEtal2008apjSpectra}
{Burrows}, A., {Budaj}, J., \& {Hubeny}, I. 2008, \apj, 678, 1436

\bibitem[{{Campo} {et~al.}(2011){Campo}, {Harrington}, {Hardy}, {Stevenson},
  {Nymeyer}, {Ragozzine}, {Lust}, {Anderson}, {Collier-Cameron}, {Blecic},
  {Britt}, {Bowman}, {Wheatley}, {Loredo}, {Deming}, {Hebb}, {Hellier},
  {Maxted}, {Pollaco}, \& {West}}]{CampoEtal2011apjWASP12b}
{Campo}, C.~J., {Harrington}, J., {Hardy}, R.~A., {et~al.} 2011, \apj, 727, 125

\bibitem[{{Castelli} \& {Kurucz}(2004)}]{Castelli2004}
{Castelli}, F. \& {Kurucz}, R.~L. 2004, arXiv:astro-ph/0405087

\bibitem[{{Charbonneau} {et~al.}(2005){Charbonneau}, {Allen}, {Megeath},
  {Torres}, {Alonso}, {Brown}, {Gilliland}, {Latham}, {Mandushev}, {O'Donovan},
  \& {Sozzetti}}]{CharbonneauEtal2005apjTrES1}
{Charbonneau}, D., {Allen}, L.~E., {Megeath}, S.~T., {et~al.} 2005, \apj, 626,
  523

\bibitem[{{Christiansen} {et~al.}(2010){Christiansen}, {Ballard},
  {Charbonneau}, {Madhusudhan}, {Seager}, {Holman}, {Wellnitz}, {Deming},
  {A'Hearn}, \& {the EPOXI Team}}]{CristiansenEtal2010ApJ-HATP7b}
{Christiansen}, J.~L., {Ballard}, S., {Charbonneau}, D., {et~al.} 2010, \apj,
  710, 97

\bibitem[{{Collier Cameron} {et~al.}(2006){Collier Cameron}, {Pollacco},
  {Street}, {Lister}, {West}, {Wilson}, {Pont}, {Christian}, {Clarkson},
  {Enoch}, {Evans}, {Fitzsimmons}, {Haswell}, {Hellier}, {Hodgkin}, {Horne},
  {Irwin}, {Kane}, {Keenan}, {Norton}, {Parley}, {Osborne}, {Ryans}, {Skillen},
  \& {Wheatley}}]{Cameron2006-SuperWASP}
{Collier Cameron}, A., {Pollacco}, D., {Street}, R.~A., {et~al.} 2006, \mnras,
  373, 799

\bibitem[{{Collier Cameron} {et~al.}(2007){Collier Cameron}, {Wilson}, {West},
  {Hebb}, {Wang}, {Aigrain}, {Bouchy}, {Christian}, {Clarkson}, {Enoch},
  {Esposito}, {Guenther}, {Haswell}, {H{\'e}brard}, {Hellier}, {Horne},
  {Irwin}, {Kane}, {Loeillet}, {Lister}, {Maxted}, {Mayor}, {Moutou}, {Parley},
  {Pollacco}, {Pont}, {Queloz}, {Ryans}, {Skillen}, {Street}, {Udry}, \&
  {Wheatley}}]{Cameron2007-SuperWASP}
{Collier Cameron}, A., {Wilson}, D.~M., {West}, R.~G., {et~al.} 2007, \mnras,
  380, 1230

\bibitem[{{Deleuil} {et~al.}(2008){Deleuil}, {Deeg}, {Alonso}, {Bouchy},
  {Rouan}, {Auvergne}, {Baglin}, {Aigrain}, {Almenara}, {Barbieri}, {Barge},
  {Bruntt}, {Bord{\'e}}, {Collier Cameron}, {Csizmadia}, {de La Reza},
  {Dvorak}, {Erikson}, {Fridlund}, {Gandolfi}, {Gillon}, {Guenther}, {Guillot},
  {Hatzes}, {H{\'e}brard}, {Jorda}, {Lammer}, {L{\'e}ger}, {Llebaria},
  {Loeillet}, {Mayor}, {Mazeh}, {Moutou}, {Ollivier}, {P{\"a}tzold}, {Pont},
  {Queloz}, {Rauer}, {Schneider}, {Shporer}, {Wuchterl}, \&
  {Zucker}}]{Deleuil2008-CoRoT3b}
{Deleuil}, M., {Deeg}, H.~J., {Alonso}, R., {et~al.} 2008, \aap, 491, 889

\bibitem[{{D{\'e}sert} {et~al.}(2009){D{\'e}sert}, {Lecavelier des Etangs},
  {H{\'e}brard}, {Sing}, {Ehrenreich}, {Ferlet}, \&
  {Vidal-Madjar}}]{Desert2009}
{D{\'e}sert}, J.-M., {Lecavelier des Etangs}, A., {H{\'e}brard}, G., {et~al.}
  2009, \apj, 699, 478

\bibitem[{{Eastman} {et~al.}(2010){Eastman}, {Siverd}, \&
  {Gaudi}}]{EastmanEtal2010apjLeapSec}
{Eastman}, J., {Siverd}, R., \& {Gaudi}, B.~S. 2010, \pasp, 122, 935

\bibitem[{{Fazio} {et~al.}(2004)}]{Fazio2004IRAC}
{Fazio}, G.~G. {et~al.} 2004, Astrophy. J. Suppl. Ser., 154, 10

\bibitem[{{Fortney} {et~al.}(2008){Fortney}, {Lodders}, {Marley}, \&
  {Freedman}}]{Fortney2008}
{Fortney}, J.~J., {Lodders}, K., {Marley}, M.~S., \& {Freedman}, R.~S. 2008,
  \apj, 678, 1419

\bibitem[{{Fortney} {et~al.}(2007){Fortney}, {Marley}, \&
  {Barnes}}]{FortneyEtal2007apjPlanetRadii}
{Fortney}, J.~J., {Marley}, M.~S., \& {Barnes}, J.~W. 2007, \apj, 659, 1661

\bibitem[{{Freedman} {et~al.}(2008){Freedman}, {Marley}, \&
  {Lodders}}]{Freedman08}
{Freedman}, R.~S., {Marley}, M.~S., \& {Lodders}, K. 2008, \apjs, 174, 504

\bibitem[{{Fressin} {et~al.}(2010){Fressin}, {Knutson}, {Charbonneau},
  {O'Donovan}, {Burrows}, {Deming}, {Mandushev}, \&
  {Spiegel}}]{FressinEtal2010ApJ-Tres3}
{Fressin}, F., {Knutson}, H.~A., {Charbonneau}, D., {et~al.} 2010, \apj, 711,
  374

\bibitem[{Gelman(2002)}]{Gelman2002}
Gelman, A. 2002, in Encyclopedia of Environmetrics vol 3, ed. A.~H. E.-S. .
  W.~W. Piegorsch (Chichester, NY: Wiley), 1634

\bibitem[{Gelman \& Rubin(1992)}]{GelmanRubin1992}
Gelman, A. \& Rubin, D. 1992, StaSc, 7, 457

\bibitem[{{Gim{\'e}nez} \& {Bastero}(1995)}]{GimenezBastero1995}
{Gim{\'e}nez}, A. \& {Bastero}, M. 1995, \apss, 226, 99

\bibitem[{{Goldreich} \& {Soter}(1966)}]{Goldreich1966}
{Goldreich}, P. \& {Soter}, S. 1966, Icar, 5, 375

\bibitem[{{Harrington} {et~al.}(2007){Harrington}, {Luszcz}, {Seager},
  {Deming}, \& {Richardson}}]{HarringtonEtal2007natHD149026b}
{Harrington}, J., {Luszcz}, S., {Seager}, S., {Deming}, D., \& {Richardson},
  L.~J. 2007, \nat, 447, 691

\bibitem[{{H{\'e}brard} {et~al.}(2008){H{\'e}brard}, {Bouchy}, {Pont},
  {Loeillet}, {Rabus}, {Bonfils}, {Moutou}, {Boisse}, {Delfosse}, {Desort},
  {Eggenberger}, {Ehrenreich}, {Forveille}, {Lagrange}, {Lovis}, {Mayor},
  {Pepe}, {Perrier}, {Queloz}, {Santos}, {S{\'e}gransan}, {Udry}, \&
  {Vidal-Madjar}}]{Hebrard2008-XO3b}
{H{\'e}brard}, G., {Bouchy}, F., {Pont}, F., {et~al.} 2008, \aap, 488, 763

\bibitem[{{Hubeny} {et~al.}(2003){Hubeny}, {Burrows}, \&
  {Sudarsky}}]{Hubeny2003}
{Hubeny}, I., {Burrows}, A., \& {Sudarsky}, D. 2003, \apj, 594, 1011

\bibitem[{{Husnoo} {et~al.}(2011){Husnoo}, {Pont}, {H{\'e}brard}, {Simpson},
  {Mazeh}, {Bouchy}, {Moutou}, {Arnold}, {Boisse}, {D{\'{\i}}az},
  {Eggenberger}, \& {Shporer}}]{Husnoo2011-WASP14b}
{Husnoo}, N., {Pont}, F., {H{\'e}brard}, G., {et~al.} 2011, \mnras, 413, 2500

\bibitem[{Jeffreys(1961)}]{Jeffreys61}
Jeffreys, H. 1961, Theory of Probability (3rd ed.; Oxford: Oxford Univ. Press)

\bibitem[{{Johns-Krull} {et~al.}(2008){Johns-Krull}, {McCullough}, {Burke},
  {Valenti}, {Janes}, {Heasley}, {Prato}, {Bissinger}, {Fleenor}, {Foote},
  {Garcia-Melendo}, {Gary}, {Howell}, {Mallia}, {Masi}, \&
  {Vanmunster}}]{Johns-Krull2008-XO-3b}
{Johns-Krull}, C.~M., {McCullough}, P.~R., {Burke}, C.~J., {et~al.} 2008, \apj,
  677, 657

\bibitem[{{Joshi} {et~al.}(2009){Joshi}, {Pollacco}, {Collier Cameron},
  {Skillen}, {Simpson}, {Steele}, {Street}, {Stempels}, {Christian}, {Hebb},
  {Bouchy}, {Gibson}, {H{\'e}brard}, {Keenan}, {Loeillet}, {Meaburn}, {Moutou},
  {Smalley}, {Todd}, {West}, {Anderson}, {Bentley}, {Enoch}, {Haswell},
  {Hellier}, {Horne}, {Irwin}, {Lister}, {McDonald}, {Maxted}, {Mayor},
  {Norton}, {Parley}, {Perrier}, {Pont}, {Queloz}, {Ryans}, {Smith}, {Udry},
  {Wheatley}, \& {Wilson}}]{Joshi2009-WASP14b}
{Joshi}, Y.~C., {Pollacco}, D., {Collier Cameron}, A., {et~al.} 2009, \mnras,
  392, 1532

\bibitem[{{Karkoschka} \& {Tomasko}(2010)}]{KarkoschkaTomasko2010}
{Karkoschka}, E. \& {Tomasko}, M.~G. 2010, Icar, 205, 674

\bibitem[{{Knutson} {et~al.}(2008){Knutson}, {Charbonneau}, {Allen}, {Burrows},
  \& {Megeath}}]{Knutson08}
{Knutson}, H.~A., {Charbonneau}, D., {Allen}, L.~E., {Burrows}, A., \&
  {Megeath}, S.~T. 2008, \apj, 673, 526

\bibitem[{{Knutson} {et~al.}(2009{\natexlab{a}}){Knutson}, {Charbonneau},
  {Burrows}, {O'Donovan}, \& {Mandushev}}]{KnutsonEtal2009ApJ-Tres4Inversion}
{Knutson}, H.~A., {Charbonneau}, D., {Burrows}, A., {O'Donovan}, F.~T., \&
  {Mandushev}, G. 2009{\natexlab{a}}, \apj, 691, 866

\bibitem[{{Knutson} {et~al.}(2009{\natexlab{b}}){Knutson}, {Charbonneau},
  {Cowan}, {Fortney}, {Showman}, {Agol}, \&
  {Henry}}]{KnutsonEtal2009apjHD149026bphase}
{Knutson}, H.~A., {Charbonneau}, D., {Cowan}, N.~B., {et~al.}
  2009{\natexlab{b}}, \apj, 703, 769

\bibitem[{{Knutson} {et~al.}(2007){Knutson}, {Charbonneau}, {Deming}, \&
  {Richardson}}]{Knutson2007}
{Knutson}, H.~A., {Charbonneau}, D., {Deming}, D., \& {Richardson}, L.~J. 2007,
  \pasp, 119, 616

\bibitem[{{Knutson} {et~al.}(2010){Knutson}, {Howard}, \&
  {Isaacson}}]{KnutsonHowardIsaacson2010ApJ-CorrStarPlanet}
{Knutson}, H.~A., {Howard}, A.~W., \& {Isaacson}, H. 2010, \apj, 720, 1569

\bibitem[{{Levenberg}(1944)}]{Levenberg1944}
{Levenberg}, K. 1944, QApMa, 2, 164

\bibitem[{{Liddle}(2007)}]{Liddle2008}
{Liddle}, A.~R. 2007, \mnras, 377, L74

\bibitem[{{Loeillet} {et~al.}(2008{\natexlab{a}}){Loeillet}, {Bouchy},
  {Deleuil}, {Royer}, {Bouret}, {Moutou}, {Barge}, {de Laverny}, {Pont},
  {Recio-Blanco}, \& {Santos}}]{Loeillet2008}
{Loeillet}, B., {Bouchy}, F., {Deleuil}, M., {et~al.} 2008{\natexlab{a}}, \aap,
  479, 865

\bibitem[{{Loeillet} {et~al.}(2008{\natexlab{b}}){Loeillet}, {Shporer},
  {Bouchy}, {Pont}, {Mazeh}, {Beuzit}, {Boisse}, {Bonfils}, {da Silva},
  {Delfosse}, {Desort}, {Ecuvillon}, {Forveille}, {Galland}, {Gallenne},
  {H{\'e}brard}, {Lagrange}, {Lovis}, {Mayor}, {Moutou}, {Pepe}, {Perrier},
  {Queloz}, {S{\'e}gransan}, {Sivan}, {Santos}, {Tsodikovich}, {Udry}, \&
  {Vidal-Madjar}}]{Loeillet2008-HATP-2b}
{Loeillet}, B., {Shporer}, A., {Bouchy}, F., {et~al.} 2008{\natexlab{b}}, \aap,
  481, 529

\bibitem[{{Lust} {et~al.}(2013){Lust}, {Britt}, {Harrington}, {Nymeyer},
  {Stevenson}, {Lust}, {Bowman}, \& {Fraine}}]{LustEtal2013apjCentering}
{Lust}, N.~B., {Britt}, D.~T., {Harrington}, J., {et~al.} 2013, \pasp,
  submitted

\bibitem[{{Machalek} {et~al.}(2008){Machalek}, {McCullough}, {Burke},
  {Valenti}, {Burrows}, \& {Hora}}]{Machalek2008-XO-1b}
{Machalek}, P., {McCullough}, P.~R., {Burke}, C.~J., {et~al.} 2008, \apj, 684,
  1427

\bibitem[{{Machalek} {et~al.}(2009){Machalek}, {McCullough}, {Burrows},
  {Burke}, {Hora}, \& {Johns-Krull}}]{MachalekEtal2009ApJ-XO2b}
{Machalek}, P., {McCullough}, P.~R., {Burrows}, A., {et~al.} 2009, \apj, 701,
  514

\bibitem[{{Madhusudhan}(2012)}]{Madhusudhan2012}
{Madhusudhan}, N. 2012, \apj, 758, 36

\bibitem[{{Madhusudhan} {et~al.}(2011{\natexlab{a}}){Madhusudhan},
  {Harrington}, {Stevenson}, {Nymeyer}, {Campo}, {Wheatley}, {Deming},
  {Blecic}, {Hardy}, {Lust}, {Anderson}, {Collier-Cameron}, {Britt}, {Bowman},
  {Hebb}, {Hellier}, {Maxted}, {Pollacco}, \&
  {West}}]{MadhusudhanEtal2011natWASP12batm}
{Madhusudhan}, N., {Harrington}, J., {Stevenson}, K.~B., {et~al.}
  2011{\natexlab{a}}, \nat, 469, 64

\bibitem[{{Madhusudhan} {et~al.}(2011{\natexlab{b}}){Madhusudhan}, {Mousis},
  {Johnson}, \& {Lunine}}]{Madhusudhan2011b}
{Madhusudhan}, N., {Mousis}, O., {Johnson}, T.~V., \& {Lunine}, J.~I.
  2011{\natexlab{b}}, \apj, 743, 191

\bibitem[{{Madhusudhan} \& {Seager}(2009)}]{MadhusudhanSeager2009}
{Madhusudhan}, N. \& {Seager}, S. 2009, \apj, 707, 24

\bibitem[{{Madhusudhan} \& {Seager}(2010)}]{MadhusudhanSeager2010}
{Madhusudhan}, N. \& {Seager}, S. 2010, \apj, 725, 261

\bibitem[{{Mandel} \& {Agol}(2002)}]{MandelAgol2002ApJtransits}
{Mandel}, K. \& {Agol}, E. 2002, \apjl, 580, L171

\bibitem[{Marquardt(1963)}]{Marquardt1963}
Marquardt, D.~W. 1963, SJAM, 11, 431

\bibitem[{{Maxted} {et~al.}(2011){Maxted}, {Anderson}, {Collier Cameron},
  {Hellier}, {Queloz}, {Smalley}, {Street}, {Triaud}, {West}, {Gillon},
  {Lister}, {Pepe}, {Pollacco}, {S{\'e}gransan}, {Smith}, \&
  {Udry}}]{Maxted2011-WASP41}
{Maxted}, P.~F.~L., {Anderson}, D.~R., {Collier Cameron}, A., {et~al.} 2011,
  \pasp, 123, 547

\bibitem[{{Morales-Calder{\'o}n} {et~al.}(2006){Morales-Calder{\'o}n},
  {Stauffer}, {Kirkpatrick}, {Carey}, {Gelino}, {Barrado y Navascu{\'e}s},
  {Rebull}, {Lowrance}, {Marley}, {Charbonneau}, {Patten}, {Megeath}, \&
  {Buzasi}}]{Morales-Calderon2006}
{Morales-Calder{\'o}n}, M., {Stauffer}, J.~R., {Kirkpatrick}, J.~D., {et~al.}
  2006, \apj, 653, 1454

\bibitem[{{Nymeyer} {et~al.}(2011){Nymeyer}, {Harrington}, {Hardy},
  {Stevenson}, {Campo}, {Madhusudhan}, {Collier-Cameron}, {Loredo}, {Blecic},
  {Bowman}, {Britt}, {Cubillos}, {Hellier}, {Gillon}, {Maxted}, {Hebb},
  {Wheatley}, {Pollacco}, \& {Anderson}}]{NymeyerEtal2011apjWASP18b}
{Nymeyer}, S., {Harrington}, J., {Hardy}, R.~A., {et~al.} 2011, \apj, 742, 35

\bibitem[{{O'Donovan} {et~al.}(2010){O'Donovan}, {Charbonneau}, {Harrington},
  {Madhusudhan}, {Seager}, {Deming}, \&
  {Knutson}}]{O'DonovanEtal2010ApJ-SpitzerTres2}
{O'Donovan}, F.~T., {Charbonneau}, D., {Harrington}, J., {et~al.} 2010, \apj,
  710, 1551

\bibitem[{{Poddan{\'y}} {et~al.}(2010){Poddan{\'y}}, {Br{\'a}t}, \&
  {Pejcha}}]{Poddany2010}
{Poddan{\'y}}, S., {Br{\'a}t}, L., \& {Pejcha}, O. 2010, 15, 297

\bibitem[{{Pollacco} {et~al.}(2006){Pollacco}, {Skillen}, {Collier Cameron},
  {Christian}, {Irwin}, {Lister}, {Street}, {West}, {Clarkson}, {Evans},
  {Fitzsimmons}, {Haswell}, {Hellier}, {Hodgkin}, {Horne}, {Jones}, {Kane},
  {Keenan}, {Norton}, {Osborne}, {Ryans}, \& {Wheatley}}]{Pollocco06}
{Pollacco}, D., {Skillen}, I., {Collier Cameron}, A., {et~al.} 2006, \apss,
  304, 253

\bibitem[{{Pont} {et~al.}(2011){Pont}, {Husnoo}, {Mazeh}, \&
  {Fabrycky}}]{Pont2011-mass-period}
{Pont}, F., {Husnoo}, N., {Mazeh}, T., \& {Fabrycky}, D. 2011, \mnras, 414,
  1278

\bibitem[{{Pont} {et~al.}(2006){Pont}, {Zucker}, \& {Queloz}}]{pont:2006}
{Pont}, F., {Zucker}, S., \& {Queloz}, D. 2006, \mnras, 373, 231

\bibitem[{{Press} {et~al.}(1992){Press}, {Teukolsky}, {Vetterling}, \&
  {Flannery}}]{PressEtalNumRec}
{Press}, W.~H., {Teukolsky}, S.~A., {Vetterling}, W.~T., \& {Flannery}, B.~P.
  1992, {Numerical Recipes in FORTRAN. The Art of Scientific Computing} (2nd
  ed.; Cambridge, Cambridge Univ. Press)

\bibitem[{{Raftery}(1995)}]{Raftery1995-BIC}
{Raftery}, A.~E. 1995, Sociological Mehodology, 25, 111

\bibitem[{{Reach} {et~al.}(2005){Reach}, {Megeath}, {Cohen}, {Hora}, {Carey},
  {Surace}, {Willner}, {Barmby}, {Wilson}, {Glaccum}, {Lowrance}, {Marengo}, \&
  {Fazio}}]{Reach2005-IRACCalibration}
{Reach}, W.~T., {Megeath}, S.~T., {Cohen}, M., {et~al.} 2005, \pasp, 117, 978

\bibitem[{{Rothman} {et~al.}(2005){Rothman}, {Jacquemart}, {Barbe}, {Chris
  Benner}, {Birk}, {Brown}, {Carleer}, {Chackerian}, {Chance}, {Coudert},
  {Dana}, {Devi}, {Flaud}, {Gamache}, {Goldman}, {Hartmann}, {Jucks}, {Maki},
  {Mandin}, {Massie}, {Orphal}, {Perrin}, {Rinsland}, {Smith}, {Tennyson},
  {Tolchenov}, {Toth}, {Vander Auwera}, {Varanasi}, \&
  {Wagner}}]{Rothman2005-HITRAN}
{Rothman}, L.~S., {Jacquemart}, D., {Barbe}, A., {et~al.} 2005, \jqsrt, 96, 139

\bibitem[{{Seager} \& {Deming}(2010)}]{SeagerDeming2010}
{Seager}, S. \& {Deming}, D. 2010, \araa, 48, 631

\bibitem[{Sivia \& Skilling(2006)}]{Sivia06}
Sivia, D.~S. \& Skilling, J. 2006, Data Analysis: A Bayesian Tutorial (2nd ed.;
  Oxford: Oxford Univ. Press)

\bibitem[{{Spiegel} {et~al.}(2009){Spiegel}, {Silverio}, \&
  {Burrows}}]{Spiegel2009}
{Spiegel}, D.~S., {Silverio}, K., \& {Burrows}, A. 2009, \apj, 699, 1487

\bibitem[{{Stevenson} {et~al.}(2012{\natexlab{a}}){Stevenson}, {Harrington},
  {Fortney}, {Loredo}, {Hardy}, {Nymeyer}, {Bowman}, {Cubillos}, {Bowman}, \&
  {Hardin}}]{StevensonEtal2012apjHD149026b}
{Stevenson}, K.~B., {Harrington}, J., {Fortney}, J.~J., {et~al.}
  2012{\natexlab{a}}, \apj, 754, 136

\bibitem[{{Stevenson} {et~al.}(2012{\natexlab{b}}){Stevenson}, {Harrington},
  {Lust}, {Lewis}, {Montagnier}, {Moses}, {Visscher}, {Blecic}, {Hardy},
  {Cubillos}, \& {Campo}}]{StevensonEtal2012apjGJ436c}
{Stevenson}, K.~B., {Harrington}, J., {Lust}, N.~B., {et~al.}
  2012{\natexlab{b}}, \apj, 755, 9

\bibitem[{{Stevenson} {et~al.}(2010){Stevenson}, {Harrington}, {Nymeyer},
  {Madhusudhan}, {Seager}, {Bowman}, {Hardy}, {Deming}, {Rauscher}, \&
  {Lust}}]{StevensonEtal2010Natur}
{Stevenson}, K.~B., {Harrington}, J., {Nymeyer}, S., {et~al.} 2010, \nat, 464,
  1161

\bibitem[{{Tinetti} {et~al.}(2007){Tinetti}, {Vidal-Madjar}, {Liang},
  {Beaulieu}, {Yung}, {Carey}, {Barber}, {Tennyson}, {Ribas}, {Allard},
  {Ballester}, {Sing}, \& {Selsis}}]{Tinetti2007Nature}
{Tinetti}, G., {Vidal-Madjar}, A., {Liang}, M.-C., {et~al.} 2007, \nat, 448,
  169

\bibitem[{{Triaud} {et~al.}(2009){Triaud}, {Queloz}, {Bouchy}, {Moutou},
  {Collier Cameron}, {Claret}, {Barge}, {Benz}, {Deleuil}, {Guillot},
  {H{\'e}brard}, {Lecavelier Des {\'E}tangs}, {Lovis}, {Mayor}, {Pepe}, \&
  {Udry}}]{Triaud2009-CoRoT3b}
{Triaud}, A.~H.~M.~J., {Queloz}, D., {Bouchy}, F., {et~al.} 2009, \aap, 506,
  377

\bibitem[{{Werner} {et~al.}(2004)}]{Werner2004}
{Werner}, M.~W. {et~al.} 2004, \apjs, 154, 1

\bibitem[{{Winn} {et~al.}(2008){Winn}, {Holman}, {Torres}, {McCullough},
  {Johns-Krull}, {Latham}, {Shporer}, {Mazeh}, {Garcia-Melendo}, {Foote},
  {Esquerdo}, \& {Everett}}]{winn:2008}
{Winn}, J.~N., {Holman}, M.~J., {Torres}, G., {et~al.} 2008, \apj, 683, 1076

\bibitem[{{Winn} {et~al.}(2007){Winn}, {Johnson}, {Peek}, {Marcy}, {Bakos},
  {Enya}, {Narita}, {Suto}, {Turner}, \& {Vogt}}]{Winn2007-HATP2b}
{Winn}, J.~N., {Johnson}, J.~A., {Peek}, K.~M.~G., {et~al.} 2007, \apjl, 665,
  L167

\end{thebibliography}

\end{document}